# AAA: an Adaptive Mechanism for Locally Differentially Private Mean Estimation

Fei Wei
National University of Singapore
feiwei@nus.edu.sg

Ergute Bao
National University of Singapore
ergute@comp.nus.edu.sg

Xiaokui Xiao
National University of Singapore
xkxiao@nus.edu.sg

Yin Yang
Hamad Bin Khalifa University
yyang@hbku.edu.qa

Bolin Ding
Alibaba Group
bolin.ding@alibaba-inc.com

## ABSTRACT

*Local differential privacy* (*LDP*) is a strong privacy standard that has been adopted by popular software systems, including Chrome, iOS, MacOS, and Windows. The main idea is that each individual perturbs their own data locally, and only submits the resulting noisy version to a data aggregator. Although much effort has been devoted to computing various types of aggregates and building machine learning applications under LDP, research on fundamental perturbation mechanisms has not achieved significant improvement in recent years. Towards a more refined result utility, existing works in the literature mainly focus on improving the *worst-case* guarantee. However, this approach does not necessarily promise a better *average* performance given the fact that the data in practice obey a certain distribution, which is not known beforehand.

In this paper, we propose the *advanced adaptive additive* (*AAA*) mechanism, which is a distribution-aware approach that addresses the average utility and tackles the classical *mean estimation* problem. AAA is carried out in a two-step approach: first, as the global data distribution is not available beforehand, the data aggregator selects a random subset of individuals to compute a (noisy) quantized data descriptor; then, in the second step, the data aggregator collects data from the remaining individuals, which are perturbed in a distribution-aware fashion. The perturbation involved in the latter step is obtained by solving an optimization problem, which is formulated with the data descriptor obtained in the former step and the desired properties of task-determined utilities. We provide rigorous privacy proofs and utility analyses, as well as extensive experiments comparing AAA with state-of-the-art mechanisms. The evaluation results demonstrate that the AAA mechanism consistently outperforms existing solutions with a clear margin in terms of result utility, on a wide range of privacy constraints and real-world and synthetic datasets.

## 1 INTRODUCTION

*Differential privacy* (*DP*) [20] is a strong and mathematically rigorous metric that evaluates the privacy guarantee provided by randomization-based data-releasing mechanisms. The concept of DP was first proposed for a centralized setting involving a trusted data curator, which releases information (e.g., mean and heavy hitters) derived from an underlying dataset consisting of sensitive records. In such a setup, DP protects individuals' privacy by requiring that the data curator randomly perturbs the released information. In particular, the perturbation is carefully calibrated to ensure that using the released noisy results, an adversary equipped with arbitrary background knowledge can only have limited confidence when inferring about the underlying sensitive individual data records. Note that this centralized setting requires a trusted data curator, who has direct access to all sensitive records in the dataset. This might not be suitable for scenarios in which individuals do not want to expose their sensitive data to any party, including such a data curator. Moreover, even if the data curator itself can be trusted, it still faces the burden of protecting sensitive data against malicious intruders. Failure to do so may lead to data breaches (e.g., in the recent incidents involving Optus [1] and Samsung [2]), which violate the individuals' privacy and damage the reputation of the data curator.

*Local differential privacy* (*LDP*) [18, 23, 35] applies the notion of DP to a different setting, in which each individual (i.e., data owner) perturbs her data *locally*, and only submits the perturbed version of her data to an untrusted *data curator*. Therefore, the data curator only collects data that is already perturbed to satisfy the rigorous LDP requirements, and malicious parties might be less incentivized to intrude into such a data curator and steal the (less sensitive) randomized dataset. Perhaps for these reasons, LDP has gained adoption rapidly ever since its proposal. In particular, LDP has been applied in common software systems that collect usage information, which include Apple iOS and MacOS [4, 43], Microsoft Windows [16], and Google Chrome [22]. LDP has also found applications in online services such as Facebook (for gathering users' behavioral data for advertisement placements) [37], and Amazon (for collecting users' shopping preferences) [41].

Driven by its successful applications, LDP has attracted much research attention in recent years. The majority of LDP-related papers focus on computing various types of statistics from the collected perturbed data, with the goal of maximizing result utility while satisfying LDP for each participating individual. These include range query [12, 14, 46], joint distribution [24, 51, 52] and marginal distribution estimation [13, 53], frequency/histogram [2, 6, 15, 36, 45], heavy hitters [5, 10, 38, 47], etc. We note that there is no single solution that consistently achieves high utility for all the tasks, as the utility varies according to different tasks.

In this paper, we focus on the fundamental LDP task of *mean estimation*, for which unbiasedness and minimal variance are desired utilities (see the detailed problem formulations in Section 3). Note that within this scope (i.e. mean estimation), there is only a narrow selection of works, including Duchi's mechanism [19], piecewise and hybrid mechanisms [44], that achieve consistently higher result utility than the classic LDP mechanisms, such as Laplace [20] and randomized response [30, 48]. For achieving better result utility, these works mainly focus on improving the *worst-case* utility guarantee. However, this approach does not necessarily provide a better *average* utility. For example, given an input data distribution dense in the *sub-optimal* regime of the mechanism, the average utility would be less ideal, even with a better worst-case guarantee. When this is true, it is natural to consider a distribution-aware approach that improves the *average* performance, instead of the overly pessimistic worst-case guarantee. This observation has not drawn sufficient attention in the literature, and motivates this work.

**Our contributions:** We propose the *advanced adaptive additive* (*AAA*) mechanism that is *adaptive* to the global data distribution while aiming to obtain high average result utility under strict LDP constraints. Achieving such a goal, however, is challenging since (i) the metric of utility varies according to the task, (ii) in the LDP setting, the global data distribution is not known beforehand, (iii) even when the data distribution is available, formulating and solving the *distribution-aware data perturbation* problem is still highly non-trivial, as elaborated later in Sections 3 and 4.

The proposed AAA mechanism addresses these challenges through a two-phase approach. In the first phase, the data curator randomly selects a subset of all participating individuals to estimate (a quantized representation of) the global data distribution, while LDP is preserved for every individual. In the second phase, each of the remaining individuals submits data perturbed by a calibrated mechanism. In particular, the perturbation in the second phase is carefully calibrated with the estimated global data distribution, which involves solving a convex optimization problem formulated for maximizing utility while enforcing LDP.

The most notable property of the perturbation noise in the second phase of AAA is that the distribution of the noise applied on the individual side depends on the value of her sensitive data. As a consequence, the utility with respect to a single individual varies as the input value varies. In particular, the more frequent data points are perturbed with smaller noises whereas the others are perturbed with larger noises, which is aligned with our goal of optimizing the average-case utility with respect to a data distribution.

Figure 1 shows an example of conditional injected noise used in AAA that is tailored for a specific input distribution. This differs from the classic approaches such as the Laplace mechanism, where the added noise is independent of the input. More detailed results and explanations on AAA will be given in Section 4.

We provide rigorous analyses for the AAA mechanism. To complement our theoretical analysis, we also conduct extensive experiments using both real and synthetic data, demonstrating that AAA consistently outperforms existing solutions by a wide margin. The rest of this paper is organized as follows. Section 2 provides the necessary background on LDP definitions as well as existing LDP mechanisms. Section 3 formally defines the problem of distribution-aware data perturbation under LDP, and Section 4 presents the proposed AAA mechanism for this purpose. Section 5 establishes the privacy guarantees and the utility analysis of AAA. Section 6 discuss the related works and Section 7 presents the experimental evaluation results. Section 8 concludes the paper with directions for future work.

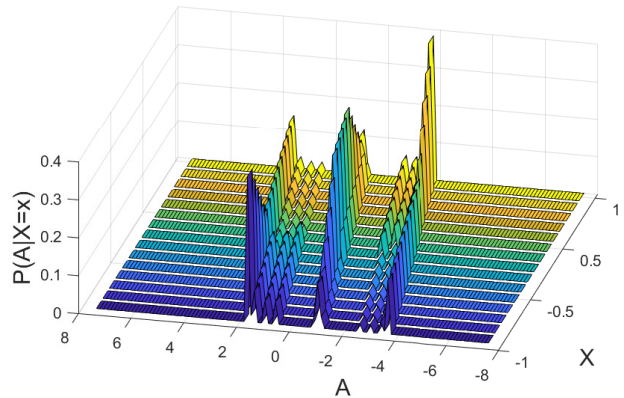

**Figure 1: The conditional probability $P_{A|X}$ that characterizes the proposed AAA mechanism, which preserves $2$-LDP and is optimized for a specific global data distribution (truncated Gaussian $N(0, 0.2^2)$ on $\mathcal{X} = [-1, 1]$). Each horizontal banner plots a probability density function of the additive noise random variable $A$ conditioning on $x \in \mathcal{X}$.**

## 2 BACKGROUND

### 2.1 Preliminaries

We consider the following scenario: there is an untrusted data curator and a set of individual clients (i.e., data owners), where the curator wishes to collect data from all clients. For example, the Census Bureau, as a data curator, may want to survey the average annual income of the entire adult population, who are the clients. Following common practice in the LDP literature, we assume that all parties are honest but curious, i.e., each party strictly follows the protocol, and at the same time tries to infer sensitive information from other parties. For simplicity, we assume that each client holds a single (1-dimensional) numeric data item $x$ supported by a finite interval $\mathcal{X} = [-\beta, \beta]$, where $\beta \in \mathbb{R}_+$.

**Caveat.** The above assumption can be made without loss of generality, as we can always rescale the values into a finite interval. However, this is not always straightforward to do, as the data items may be taken from a large or even unbounded domain (e.g., real numbers representing income data), since the existence of outliers may lead to a rescaled distribution that is only dense in some small intervals. In such cases, more sophisticated approaches such as trimming and clipping [1, 11] might be needed.

**Multi-dimensional data.** In the case that each client holds multi-dimensional data, i.e., a tuple of attributes $\mathbf{x} = (x^{(1)}, \ldots, x^{(d)})$ where each data entry is from $\mathcal{X}$, one could apply the standard trick in [44] that each client proceeds with a randomly chosen data entry, which reduces the problem to collecting one-dimensional data. This is (roughly) equivalent to randomly partitioning all clients into $d$ non-overlapping subsets, each of which reports one of the $d$ attributes. Detailed algorithm description and evaluation results are provided in the full version [49]. In the rest of the paper, we focus on the one-dimensional case where each client holds a single numeric data item.

Following existing work [19, 44], we focus on the fundamental problem of *mean estimation*, and the goal is to minimize the error of the estimated mean computed at the data curator under local differential privacy constraints. Such mechanisms typically consist of two stages: i) *Data Perturbation*, i.e., each client $i$ perturbs her local data (e.g., by injecting additive noise) before uploading it to the curator to protect sensitive information, and ii) *Data Aggregation*, i.e., the curator estimate the mean from the collected data. The perturbation can be represented by a stochastic mapping $\mathcal{M} : \mathcal{X} \to \mathcal{Y}$, which can also be seen as a *conditional distribution* $P_{Y|X}$.

In terms of the utility requirement for the perturbation mechanism, we want (i) the output to remain unbiased, that is, the expectation of the output and input needs to remain identical, and (ii) a smaller output variance to minimize the error in practice. These requirements form the utility metric of mean estimation tasks, which will play an important role in subsequent discussions. Meanwhile, regarding the privacy metric, we focus on the standard definition of pure $\epsilon$-*local differential privacy* (LDP) [18].

**Definition 1.** *For any non-negative $\epsilon$, a perturbation mechanism $\mathcal{M}$ is said to satisfy $\epsilon$-local differential privacy (LDP) if for any inputs $x \neq x' \in \mathcal{X}$ and subset $\mathcal{U} \subseteq \mathcal{Y}$, it holds that*

$$\Pr[\mathcal{M}(x) \in \mathcal{U}] \leq e^{\epsilon} \cdot \Pr[\mathcal{M}(x') \in \mathcal{U}].$$

The level of privacy protection is quantified by the parameter $\epsilon$. Roughly speaking, a smaller $\epsilon$ makes it more difficult for the adversary (including the untrusted data curator) to infer the exact input value given the perturbed version, and vice versa.

## 2.2 Data Perturbation Mechanisms under LDP

**Classic Approaches.** There are several classical data perturbation mechanisms in the literature that can be used to enforce LDP. First, we present the Laplace mechanism, assuming that each client's data item is within the interval $[-\beta, \beta]$ as described in the previous subsection.

**Lemma 1 (Laplace Mechanism [20]).** *The Laplace mechanism that injects Laplace noise of location parameter 0 and scale parameter $\frac{2\beta}{\epsilon}$ into $x$ satisfies $\epsilon$-local differential privacy, where the probability density function of the Laplace noise of parameters 0 and $\frac{2\beta}{\epsilon}$ is defined as*

$$f(x) = \frac{\epsilon}{4\beta} \exp\left(-\frac{\epsilon |x|}{2\beta}\right).$$

In case $X$ is not continuous but binary, i.e., $\mathcal{X} = \{-\beta, \beta\}$, one can apply the classic *randomized response* mechanism to enforce LDP.

**Lemma 2 (Randomized Response [30, 48]).** *Let $x \in \{-\beta, \beta\}$ be the sensitive attribute to protect. The randomized response mechanism that randomly flips the sign of $x$ satisfies $\epsilon$-local differential privacy, if the probability of flipping is as follows*

$$p_{Y|X}(y \mid x) = \begin{cases} \frac{e^{\epsilon}}{1+e^{\epsilon}}, & \text{if } y = x, \\ \frac{1}{1+e^{\epsilon}}, & \text{otherwise.} \end{cases}$$

The original randomized response mechanism is only discussed for the case where the client's data is binary. For mean estimation over $\mathcal{X} = [-\beta, \beta]$, an extension of the randomized response mechanism proposed by Duchi et al. [19], presented next.

**Duchi et al.'s Mechanism.** Duchi et al. propose a mechanism to perturb data under LDP [19]. Specifically, the algorithm takes as input $X \in [-\beta, \beta]$ and outputs a binary $Y \in \left\{ -\frac{e^{\epsilon}+1}{e^{\epsilon}-1} \cdot \beta, \frac{e^{\epsilon}+1}{e^{\epsilon}-1} \cdot \beta \right\}$ with probabilities

$$p_{Y|X}(y \mid x) = \begin{cases} \frac{e^{\epsilon}-1}{2e^{\epsilon}+2} \cdot \frac{x}{\beta} + \frac{1}{2}, & \text{if } y = \frac{e^{\epsilon}+1}{e^{\epsilon}-1} \cdot \beta, \\ -\frac{e^{\epsilon}-1}{2e^{\epsilon}+2} \cdot \frac{x}{\beta} + \frac{1}{2}, & \text{if } y = -\frac{e^{\epsilon}+1}{e^{\epsilon}-1} \cdot \beta. \end{cases}$$

It is easy to verify that $Y$ is an unbiased estimator of $X$ with variance

$$\text{Var}[Y \mid X = x] = \beta^2 \cdot \left(\frac{e^{\epsilon}+1}{e^{\epsilon}-1}\right)^2 - x^2 \leq \beta^2 \cdot \left(\frac{e^{\epsilon}+1}{e^{\epsilon}-1}\right)^2. \quad (1)$$

**Piecewise Mechanism and Hybrid Mechanism.** Wang et al. propose the piecewise mechanism (PM) [44]. PM takes as input some data $X \in [-1, 1]$. For a more general case that $X' \in [-\beta, \beta]$, [44] shows that we can scale it to $X \in [-1, 1]$, apply PM, and return the scaled output $\beta \cdot y$. Unlike Duchi's mechanism, in PM the support of the output is not discrete but a finite interval $[-C, C]$ where $C = \frac{e^{\epsilon/2}+1}{e^{\epsilon/2}-1}$. The probability density function of the piecewise mechanism is a piecewise constant function

$$f_{Y|X}(y \mid x) := \begin{cases} p, & \text{if } y \in [l(x), r(x)], \\ \frac{p}{e^{\epsilon}}, & \text{if } y \in [-C, l(x)) \cup (r(x), C]. \end{cases}$$

Here, $p = \frac{e^{\epsilon}-e^{\epsilon/2}}{2e^{\epsilon/2}+2}$, $l(x) = \frac{C+1}{2} \cdot x - \frac{C-1}{2}$, and $r(x) = l(x) + C - 1$. Accordingly, given any numeric data $x \in [-1, 1]$, [44] proves that that $Y$ is an unbiased estimate of $X$, and the conditional variance of $Y$ is

$$\text{Var}[Y \mid X = x] = \frac{x^2}{e^{\epsilon/2}-1} + \frac{e^{\epsilon/2}+3}{3(e^{\epsilon/2}-1)^2} \leq \frac{4e^{\epsilon/2}}{3(e^{\epsilon/2}-1)^2}. \quad (2)$$

In the general case that $X \in [-\beta, \beta]$, the estimate remains unbiased, and the variance of the output would be multiplied by $\beta^2$. In the same paper [44], Wang et al. also propose a hybrid mechanism (HM) that combines Duchi's mechanism and PM above. According to the analysis in [44], HM achieves better *worst-case performance* compared to both of its underlying components, i.e., Duchi's mechanism and PM. However, as our experiments in Section 7 demonstrate, the result utility of HM is often worse than PM, on several real and synthetic datasets. This highlights the fact that high worst-case performance does not necessarily indicate high *average performance*, which is the focus of this paper.

## 3 PROBLEM FORMULATION

### 3.1 Rationale

Recall from Section 2.2 that, regarding the classical mean estimation task, the existing solutions for our problem setting perturb the output regardless of the input value (or distribution). This misses the opportunity of further improving result utility by performing input-dependent perturbation, as is done in Laplace mechanism, Randomized Response, Duchi's mechanism (which generalizes Randomized Response), PM, and HM. On the other hand, observe that when the perturbation is input-dependent, the result utility can also depend on the input value. For instance, the variance of the output in Duchi's mechanism reaches its worst case when the input $x = 0$, according to Eq. (1). This means that in the unfortunate case that most clients hold a zero (or close to zero) value, Duchi's mechanism would perform poorly. Similarly, according to Eq. (2), the worst case of PM occurs when the absolute value of $x$ is large (bounded by $\beta$ in our problem setting), meaning that PM would perform poorly in the extreme case that most clients hold a value close to either $\beta$ or $-\beta$. This issue is recognized in [44], and the authors attempt to address it with HM. However, HM focuses on optimizing worst-case performance, and its result utility is lower than PM on real data and synthetic data following common probability distributions, as shown in our experiments. This leads to the question: regarding the mean estimation task, can we design an *adaptive* LDP mechanism that works well for the average case?

The answer to the above question is clearly positive, *if each client has a certain level of knowledge regarding the global data distribution before the perturbation is performed.* For example, if all clients know that a good majority of data items are close to zero, then it is probably best to avoid Duchi's mechanism, whose worst-case utility is reached when $x = 0$, and instead apply a mechanism that promises good utility for close-to-zero inputs such as PM, according to Eq. (2). Meanwhile, observe that Duchi's mechanism, PM, and their combination HM are just three specific instances in the vast design space of LDP mechanism that performs input-dependent perturbation. In fact, even if all three methods have rather poor performance, e.g., when half of the clients have close-to-zero data values (worst case for Duchi's method) and the other half have close-to-$\beta$ data values (worst case for PM), it is still possible to design an effective data perturbation scheme for such a scenario that *exploits the knowledge of the global data distribution.*

The above reasoning motivates a *distribution-aware data perturbation* mechanism for enforcing LDP, in which the perturbation depends on not only the input value, but also an (estimation of) the input value distribution. In the following subsections, we formalize this idea and define the optimization objective and constraints.

## 3.2 Optimization Problem Formulation

Motivated by the discussion in Section 2.1 and above, we start with the following abstract problem definition, and gradually fill out the mathematical details.

Problem 1 (Optimal Mean Estimation Mechanism under LDP). *Given $\mathcal{D}$ represents the set of numeric data items $x \in \mathcal{X} = [-\beta, \beta]$ held by local clients, we want to design a mechanism $\mathcal{M} : \mathcal{X} \rightarrow \mathcal{Y}$ can be represented as a conditional probability density function $f_{Y|X}$, i.e., $Y \sim f_{Y|X}$. By Definition 1, for any valid data items $x, x' \in \mathcal{X}$, and any outputs $y$, mechanism $\mathcal{M}$ should satisfy $f_{Y|X}(y \mid x) \leq e^\epsilon \cdot f_{Y|X}(y \mid x')$.*

**Bias-free requirement.** The mean of all samples in the dataset is $\bar{X} = \frac{1}{|\mathcal{D}|} \sum_{x \in \mathcal{D}} x$. Meanwhile, with the conditional $f_{Y|X}$, the expected mean $\mathbb{E}[\bar{Y}] = \frac{1}{|\mathcal{D}|} \cdot \sum_{x \in \mathcal{D}} \mathbb{E}[Y|X = x]$. For ensuring the estimated mean to be unbiased, i.e., $\mathbb{E}[\bar{Y}] = \bar{X}$, it suffices to require $\mathbb{E}[Y \mid X = x] = x$ for all $x \in \mathcal{X}$.

**Minimized variance requirement.** Problem 1 states that we aim to minimize the total variance of $Y$, i.e. $\mathrm{Var}(Y)$. Next, we show that minimizing the total variance of $Y$ is equivalent to minimizing the expected conditional variance $\mathbb{E}[\mathrm{Var}(Y \mid X)]$. By the law of total variance [50], we have $\mathrm{Var}(Y) = \mathbb{E}[\mathrm{Var}(Y \mid X)] + \mathrm{Var}(\mathbb{E}[Y \mid X])$, where the first term is the expected variance, and the second term

$$\mathrm{Var}(\mathbb{E}[Y \mid X]) = \mathbb{E}[\mathbb{E}[Y \mid X]^2] - \mathbb{E}[\mathbb{E}[Y \mid X]]^2 = \mathrm{Var}(X).$$

The equalities hold as we require $Y$ to be bias-free, i.e., $\mathbb{E}[Y \mid X = x] = x$ for all $x \in \mathcal{X}$. Note that $\mathrm{Var}(X)$ is determined by the given data distribution and the data domain. Thus, it can be considered as a constant in the optimization problem. Accordingly, minimizing the expected conditional variance $\mathbb{E}[\mathrm{Var}(Y \mid X)]$ is equivalent to minimizing the total variance $\mathrm{Var}(Y)$.

**Optimization program.** The above discussions lead to the following optimization program:

$$
\begin{aligned}
\underset{f_{Y|X}}{\text{minimize}} \quad & \mathbb{E}[\mathrm{Var}(Y \mid X)] \\
\text{subject to} \quad & \mathbb{E}[Y \mid X = x] = x \text{ for all } x \in \mathcal{X}, \\
& f_{Y|X}(y \mid x) \leq e^\epsilon \cdot f_{Y|X}(y \mid x'), \forall x, x' \in \mathcal{X}, y \in \mathcal{Y}.
\end{aligned}
\tag{3}
$$

While the constrained optimization program above is a valid problem formulation, it is difficult to solve since (i) the global distribution of data held by local clients is not known beforehand, and (ii) the problem is defined in a continuous domain, and it is unclear how to solve for the optimal $f_{Y|X}$, which has a vast search space. To make the problem tractable, we first quantize the representation for the distribution of variable $X$, explained in detail in the next subsection.

## 3.3 Quantized Input Data Distribution

Recall from Section 2.1 that each client holds a private data item $x \in [-\beta, \beta]$, to better capture the statistical characteristic of the data collection, we provide a *randomized rounding algorithm* as follows. First, we split the domain $\mathcal{X}$ into $N$ intervals such that the width of each interval is

$$\sigma = \frac{|\beta - (-\beta)|}{N} = \frac{2\beta}{N}. \tag{4}$$

Accordingly, using the quantization parameter $\sigma$ defined above, the edges of the intervals are

$$x_j = -\beta + j \cdot \sigma \text{ for } j = 0, \ldots, N. \tag{5}$$

Next, let $r(x)$ be a randomized rounding scheme. For any sample $x \in \mathcal{D}$, assume $x$ falls into an interval $[x_i, x_{i+1}]$, we define a randomized rounding mechanism $r(x)$ that $x$ would be rounded to the nearest left index $x_i$ with probability $w_i(x)$ or the right index $x_{i+1}$ in probability

$$r(x) = \begin{cases} x_i, & \text{with probability } w_i(x), \\ x_{i+1}, & \text{with probability } w_{i+1}(x). \end{cases}$$

**Privacy requirement.** Recall from Section 2.1 that in LDP, each client $i$ applies the mechanism $\mathcal{M}$ to perturb her private data $x_i$ independently, and $\mathcal{M} : \mathcal{X} \mapsto \mathcal{Y}$ can be represented as a conditional probability density function $f_{Y|X}$, i.e., $Y \sim f_{Y|X}$. By Definition 1, for any valid data items $x, x' \in \mathcal{X}$, and any outputs $y$, mechanism $\mathcal{M}$ should satisfy $f_{Y|X}(y \mid x) \leq e^\epsilon \cdot f_{Y|X}(y \mid x')$.

Here, we define

$$w_i(x) = \begin{cases} 1 - \frac{|x - x_i|}{\sigma}, & \text{if } x \in [x_{i-1}, x_{i+1}], \\ 0, & \text{otherwise.} \end{cases} \quad (6)$$

Eq. (6) implies that $\sum_{i=0}^{N} w_i(x) = 1$ such that the randomized rounding mechanism is valid.

Thus, we derive a discrete probability mass function $p_X$ on $\{x_j : j = 0, \ldots, N\}$ by rounding all the samples in $\mathcal{D}$, and the probability of having $x_i$ is:

$$p_X(x_i) = \sum_{x \in \mathcal{D}} \Pr[r(x) = x_i | X = x] \cdot \Pr[X = x] = \frac{1}{|\mathcal{D}|} \sum_{x \in \mathcal{D}} w_i(x). \quad (7)$$

The validity of the function in Eq.(7) as a probability measure can be easily justified, as Eqs. (6) and (7) imply $\sum_{i=0}^{N} p_X(x_i) = 1$.

### 3.4 Global Data Distribution Estimation

Next, we demonstrate how to obtain a quantized distribution descriptor while enforcing LDP requirements. Clearly, this maps to an LDP-compliant discrete distribution estimation problem, which has been studied extensively in the LDP literature, e.g., in [2, 6, 15, 36, 45]. Typical solutions for locally differential private distribution estimation usually consist of two steps: 1) let all clients perturb the original data, e.g., using random-response-based perturbation algorithms and 2) the curator collects all the perturbed data and reconstructs the distribution. Note the quantization can be performed either before the perturbation or after, it mainly depends on the perturbation mechanism of choice. Existing solutions mainly differ in the perturbation strategy and the reconstruction algorithms.

As distribution estimation is not the focus of this work, one can freely apply any $\epsilon$-LDP distribution estimator. In our implementation, we use the classic randomized response mechanism, where each client in the sample set reports truthfully to one of the $N + 1$ edges with probability $\frac{\exp(\epsilon)}{\exp(\epsilon) + N}$, and randomly picks one of the rest $N$ edges otherwise. After collecting all the perturbed responses, the server reconstructs an estimate for $P_X$ by solving a linear programming problem. We provide the complete algorithm in the full paper [49].

One may note that in our solution, instead of letting all clients participate in the distribution estimation, we use a random subset of the clients for this computation, as elaborated later in Section 4.1. Lastly, with any distribution estimator, let $\hat{p}_X$ be the result discrete descriptor that estimates the quantized distribution, in the following discussions, to provide a provable performance guarantee of the result mechanism, we assume that $\hat{p}_X$ satisfies the following properties:

(1) Validity: $\hat{p}_X$ is a valid probability mass function, i.e.,

$$0 \le \hat{p}_X(x_j) \le 1 \text{ and } \sum_{j=0}^{N} \hat{p}_X(x_j) = 1.$$

(2) Bounded relative error: The relative error of the estimated probability mass associated with any index is no larger than $\psi \ge 0$, i.e.,

$$\sup_{j=0,\ldots,N} \left| \frac{p_X(x_j) - \hat{p}_X(x_j)}{p_X(x_j)} \right| \le \psi.$$

**Remark.** Beyond the discussion in Section 2, the achievability of the second condition is subject to the rescaling process (which determines the value domain and result distribution) as well as the original data distribution. When the data distribution is highly scattered, more sophisticated pre-processing techniques are needed to derive provable error bounds. This is an orthogonal problem to this paper, and we leave it as future work.

## 4 THE AAA MECHANISM

### 4.1 Solution Overview

As an adaptive mean estimation mechanism, our solution consists of two phases, *distribution estimation* and *mean estimation*.

In the first phase, a random sample set of the clients jointly participate in a quantized distribution estimation protocol that satisfies $\epsilon$-LDP, as explained in Section 3.4. We separate the clients into groups to ensure each client spends the same privacy budget throughout the whole process. For instance, one way to perform such sampling is to let each client perform a Bernoulli test with probability $s$; if the test result is positive, then the client participates in the LDP-compliant distribution estimation protocol, and vice versa. Here, the size of the sample set is decided by the ratio parameter $s$, which will be discussed more in Section 7. Note that clients in the sample set have depleted their privacy budget, and, thus, need to be excluded from the remaining procedures.

In the second phase, the remaining clients participate in the sub-protocol for mean estimation, using the proposed AAA mechanism explained in this section that solves the optimization program defined in Eq. (3) under $\epsilon$-LDP, using the estimated global data distribution descriptor $\hat{p}_X$ obtained in the first phase. However, it is still a challenging task to perform the second phase, which involves solving the difficult optimization problem defined in Eq. (3). Next, we discretize the optimization problem in Section 4.2, and then convert the problem to a solvable form in Section 4.3.

### 4.2 Discretizing the Optimization Problem

In what follows, assume we have a distribution descriptor $p_X$ (or $\hat{p}_X$ with LDP) on the discrete index set $\mathcal{X} = \{x_0, \ldots, x_N\}$ (as defined in Eq. (5)), we reformulate the problem in Eq. (3) such that it can be solved numerically.

Similar to the classic additive solutions, the proposed AAA mechanism injects random noise into each client's data value to obtain the perturbed version. Let $A$ be the added noise, which is a random variable obeying a conditional probability distribution characterized by the function $p_{A|X}$. For any $x \in \mathcal{X}$, we denote the random variable obtained by conditioning $A$ on $X = x$ as $(A \mid X = x)$ or simply $A_x$. A sample from $A_x$ is later added to the output of the query $x$ for perturbation. Accordingly, the perturbed output is $Y_x = x + A_x$. Here, let $A_x$ be supported with alphabet $\mathcal{A}$ that

$$\mathcal{A} \triangleq \{j \cdot \sigma : j \in \mathbb{Z}\}, \quad (8)$$

where $\sigma$ is the quantization constant defined previously in (4). To ensure the estimate remains unbiased, it suffices to require that

$$\mathbb{E}[A \mid X = x] = \sum_{a \in \mathcal{A}} a \cdot p_{A|X}(a \mid x) = 0 \quad (9)$$

hold for any $x \in \mathcal{X}$, which implies $\mathbb{E}[Y \mid X = x] = x$ and, thus, $\mathbb{E}[X] = \mathbb{E}[Y]$. While maintaining unbiasedness, our goal is to minimize the conditional variance and it holds that

$$\mathbb{E}[\text{Var}(Y \mid X)] = \sum_{x \in \mathcal{X}} \left( \sum_{a \in \mathcal{A}} a^2 \cdot p_{A|X}(a \mid x) \right) \cdot p_X(x). \quad (10)$$

See the full version [49] for a detailed derivation of Eq. (10). By Eq. (10), Problem (3) reduces to the following optimization problem.

$$\begin{aligned}
\underset{\mathbf{P}_{A|X}}{\text{minimize}} \quad & \sum_{x \in \mathcal{X}} \left( \sum_{a \in \mathcal{A}} a^2 \cdot p_{A|X}(a \mid x) \right) \cdot p_X(x) \\
\text{subject to} \quad & p_{A|X}(y - x \mid x) \le e^\epsilon \cdot p_{A|X}(y - x' \mid x') \\
& \forall x \ne x' \in \mathcal{X}, y \in \mathcal{Y}, \\
& \sum_{a \in \mathcal{A}} a \cdot p_{A|X}(a \mid x) = 0, \forall x \in \mathcal{X}.
\end{aligned} \quad (11)$$

The above discretized version of the optimization program is still rather difficult to solve, since the alphabet $\mathcal{A}$ is unbounded as defined in Eq. (8), leading to an infinite-dimensional search space. In the next subsection, we deal with this tricky problem through a novel definition of the conditional distribution $\mathbf{P}_{A|X}$, as well as a series of mathematical transformations of the optimization problem, arriving at a solvable form at the end of the section.

## 4.3 Solving the Optimization Problem

To address the problem that the alphabet $\mathcal{A}$ in Problem (11) is unbounded, we restrict the conditional distribution $\mathbf{P}_{A|X}$ to a special form that involves a finite number of unknowns. This is a main insight of the proposed AAA mechanism, and it is the key step towards transforming the optimization problem into a solvable form.

Specifically, given the quantization constant $N$ defined in Section 3, for each $x_i \in \mathcal{X}$, we choose an integer $M \ge N$, and define a tunable vector $\mathbf{Q} = (\mathbf{q}^{(0)}, \ldots, \mathbf{q}^{(N)})$ where $\mathbf{q}^{(i)} = (q_j^{(i)} \in \mathbb{R}^+ : j = -M, \ldots, M)$. Note that a valid $\mathbf{Q}$ needs to make sure that the induced vector defines a valid discrete distribution, i.e.,

$$\sum_{j=-M+1}^{M-1} q_j^{(i)} + \frac{1}{1-r} \cdot \left( q_{-M}^{(i)} + q_M^{(i)} \right) = 1 \text{ for all } i = 0 \ldots, N. \quad (12)$$

Here $r \le 1$ is a positive geometric constant (we will discuss the choice of $r$ later). Then, we define

$$p_{A|X}(a_j \mid x_i) = \begin{cases} q_{-M}^{(i)} \cdot r^{-M-j}, & \text{if } j \le -M, \\ q_j^{(i)}, & \text{if } |j| < M, \\ q_M^{(i)} \cdot r^{j-M}, & \text{if } j \ge M. \end{cases} \quad (13)$$

Intuitively, in the above definition of the conditional distribution $\mathbf{P}_{A|X}$, only the middle section involves a finite number of unknowns (i.e., elements of $\mathbf{Q}$), and the two edge sections are simply modeled by geometric series. Based on this definition, next we transform the infinite-dimensional problem defined in Eq. (11) to a finite-dimensional one.

In what follows, for simplicity, we denote $p_X(x_i)$ as $p_i$. The objective function in Eq. (11) under our setting can be restated as

$$\begin{aligned}
\underset{\mathbf{Q}}{\text{minimize}} \quad & \sum_{i=0}^{N} p_i \cdot \left( \sum_{j=-\infty}^{-M} a_j^2 \cdot q_{-M}^{(i)} r^{|j|-M} \right. \\
& \left. + \sum_{j=-M+1}^{M} a_j^2 \cdot q_j^{(i)} + \sum_{j=M}^{\infty} a_j^2 \cdot q_M^{(i)} r^{|j|-M} \right).
\end{aligned}$$

Recall that $a_j = \sigma \cdot j$, so

$$\begin{aligned}
\sum_{j=M}^{\infty} a_j^2 \cdot q_M^{(i)} r^{j-M} &= \sigma^2 \cdot q_M^{(i)} \sum_{j=0}^{\infty} (j+M)^2 r^j \\
&= \sigma^2 \cdot q_M^{(i)} \cdot \left( \frac{M^2}{1-r} + \frac{(2M-1)r}{(1-r)^2} + \frac{2r}{(1-r)^3} \right).
\end{aligned} \quad (14)$$

Please see the full version [49] for a detailed derivation of Eq. (14).

Finally, we obtain an optimization problem as follows:

$$\begin{aligned}
\underset{\mathbf{Q}}{\text{minimize}} \quad & \sum_{i=0}^{N} p_i \cdot \left[ \sum_{j=-M+1}^{M-1} a_j^2 \cdot q_j^{(i)} \right. \\
& \left. + \left( \frac{M^2}{1-r} + \frac{(2M-1)r}{(1-r)^2} + \frac{2r}{(1-r)^3} \right) \cdot \sigma^2 \left( q_{-M}^{(i)} + q_M^{(i)} \right) \right] \\
\text{subject to} \quad & q_j^{(i)} \le e^\epsilon \cdot q_{j'}^{(i')} \\
& \text{for all } (i, j) \ne (i', j') \text{ that } a_j + x_i = a_{j'} + x_{i'}, \\
& \sum_{j=-M+1}^{M-1} q_j^{(i)} + \frac{1}{1-r} \cdot \left( q_{-M}^{(i)} + q_M^{(i)} \right) = 1 \\
& \text{for all } i \in \{0, \ldots, N\}, \\
& q_j^{(i)} \ge 0, \text{ for all } j \in \{-M, \ldots, M\}.
\end{aligned} \quad (15)$$

**Cost.** The computation cost for solving the linear programming problem of Eq. (15) is independent of the number of clients (namely, the scale of data). Instead, the cost for most of the existing solvers of linear problems is in polynomial with respect to the *total number of unknowns*, which is $(N+1) \cdot (2M+1)$, a polynomial of $N$, as we set $M$ to small multiples of $N$ (see Section 7 for more details).

## 4.4 Result Perturbation Mechanism

Let $\mathbf{Q}^*$ denote the solution to the Problem (15) with respect to privacy parameters $\epsilon$. It is then fed to the algorithm to compute the conditional distribution $\mathbf{P}_{A|X}^*$ for the added noise.

Based on $\mathbf{P}_{A|X}^*$, recall that $\mathcal{A}$ is a discrete set, we define a perturbation mechanism $\mathcal{M} : \mathcal{X} \to \mathcal{Y}$ as follows. Assuming an input value $x$ fall in the interval $[x_i, x_{i+1}]$ where $x_i$ as defined as of Eq. (5),

$$\mathcal{M}(x) = \begin{cases} A_{x_i} + x_i, & \text{with probability } w_i(x) = 1 - \frac{|x - x_i|}{\sigma}, \\ A_{x_{i+1}} + x_{i+1}, & \text{otherwise.} \end{cases} \quad (16)$$

Here $A_{x_i} \sim p_{A|X}^*(a|x_i)$.

This completes the proposed AAA mechanism. We provide formal proofs and utility analyses of AAA in the next section. Figure 2 visualizes the optimized conditional noise distribution with respect

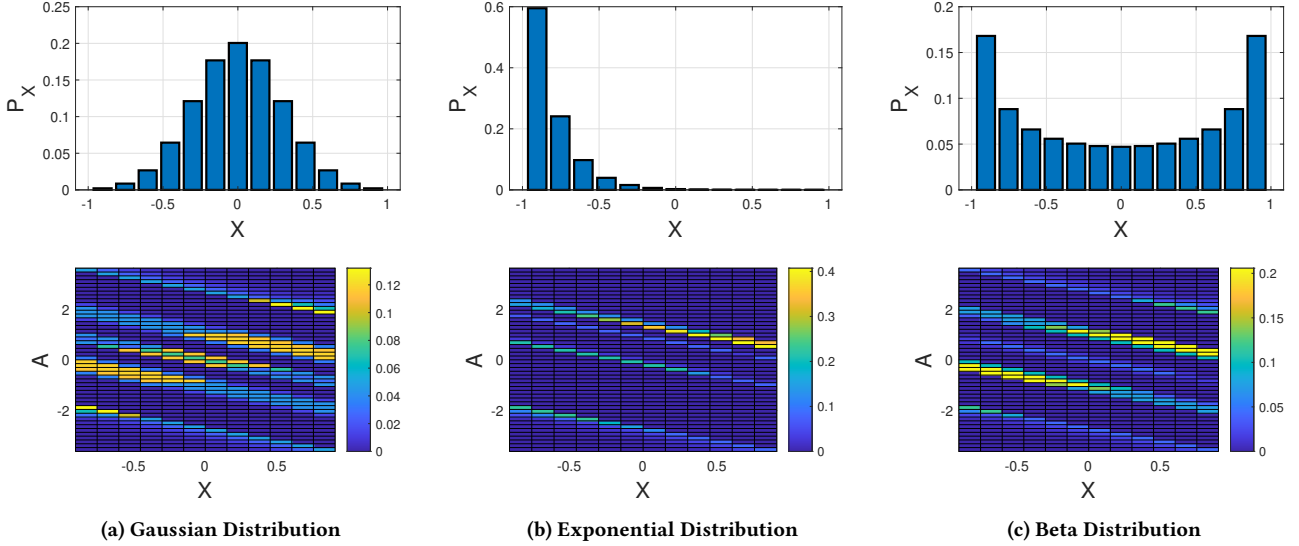

**(a) Gaussian Distribution**  **(b) Exponential Distribution**  **(c) Beta Distribution**

Figure 2: The optimized noise distribution of the AAA mechanism for several common data distributions (all truncated within the interval $[-1, 1]$). The bar plots on the top show the (quantized) data distribution for the data evaluations $X$, while the depth maps below represent the conditional distributions. Each column in the depth maps represents the discrete distribution conditioning on $X$. The brightness of the color is associated with the weight of the probability mass, as indicated by the color bar to the right of the figure. As we can see from the plots, the optimized noise distribution varies according to the data distribution.

to several common types of data distributions, in which the proposed AAA mechanism satisfies 2-LDP. As one can observe from Figure 2, in spite of the non-smoothness caused by the numerical solver, not surprisingly, given symmetric data distributions, the resulting noises are also symmetric, and vice versa. Moreover, to minimize the variance, the solver tends to concentrate the resulting noise around zero, with negative noise skew on positive valued inputs and vice versa. Meanwhile, there are probability masses on the tail ends to enforce unbiasedness, which results in an imbalanced multi-peak shape.

**Privacy enhancing after local perturbation.** A standard approach for LDP mean estimation follows the pattern of perturb-collect-aggerate, and existing LDP mechanisms mainly differ on the perturbation stage (including the proposed AAA mechanism). One can also apply privacy-enhancing techniques at other stages to improve overall privacy protection. For example, secure shuffling [7, 21] protects the individual's identity at the collecting stage by randomly reordering the perturbed data. With secure shuffling, the privacy protection of the LDP mechanism is enhanced roughly by a factor of $1/\sqrt{|\mathcal{D}|}$, where $|\mathcal{D}|$ represents the scale of the dataset (namely, the number of clients). Similarly, the clients could also aggregate their perturbed values prior to releasing them to the server using SecAgg [8]. Here, the DP guarantee for the noisy sum is provided by the sum of independent contributions of DP noises on the client side. However, in the case when some participating clients do not perturb their local data (or opt to drop off), the overall privacy guarantee degrades. In the strict LDP setting of our AAA, on the other hand, the privacy guarantee regarding any individual client is provided by his/her own perturbation, which is *independent* of the activities of other clients. Compared to the hybrid approach, such a (more) strict local DP approach might be preferable in scenarios where clients are not trustworthy.

## 5 THEORETICAL ANALYSES

### 5.1 Privacy Analysis

Before demonstrating that the perturbation of AAA satisfies $\epsilon$-LDP, we present a useful lemma.

**Lemma 3.** *Assume we have real values $a_1, a_2, a_3, a_4, w_1, w_2 \in [0, 1]$ where for any pair $(a_i, a_j)$ that $i \neq j$ we have $a_i \leq k \cdot a_j$ for $k > 1$, then it holds that*

$$w_1 \cdot a_1 + (1 - w_1) \cdot a_2 \leq k \cdot (w_2 \cdot a_3 + (1 - w_2) \cdot a_4) \,. \quad (17)$$

Lemma 3 is straightforward and can be easily proved by linearity. The complete proof is given in the full version [49].

**Claim 1.** *The perturbation mechanism $\mathcal{M}$ defined in Section 4.4 satisfies $\epsilon$-local DP.*

**Proof.** Given the definition of the mechanism $\mathcal{M}$, it is worth noting that the random choice between $A_{x_i} + x_i$ and $A_{x_{i+1}} + x_{i+1}$ in Eq. (16) can be interpreted by conditional probabilities $p(x_i \mid x) = w_i(x)$ and $p(x_{i+1} \mid x) = w_{i+1}(x)$. Meanwhile, for any output $y$, as $y = A_{x_i} + x_i$ given some $x_i$, it holds that the function $p(y \mid x_i) = p(a \mid x_i)$ for $a = y - x_i$ (as $x_i$ is considered constant here). After all, from any any inputs $x \neq x'$ to any output $y$, assuming $x \in [x_i, x_{i+1}]$ and $x' \in [x_j, x_{j+1}]$, the stochastic mapping can be characterized by

$$p_{Y|X}(y \mid x) = p_{A|X}(a \mid x_i) \cdot p(x_i \mid x) + p_{A|X}(a' \mid x_{i+1}) \cdot p(x_{i+1} \mid x),$$

and

$$p_{Y|X}(y \mid x') = p_{A|X}(\tilde{a} \mid x_j) \cdot p(x_j \mid x) + p_{A|X}(\tilde{a}' \mid x_{j+1}) \cdot p(x_{i+1} \mid x)$$

where $a + x_i = a' + x_{i+1} = \tilde{a} + x_j = \tilde{a}' + x_{j+1} = y$. Also, by our definitions, we have $p(x_i \mid x) = w_i = 1 - w_{i+1} = 1 - p(x_{i+1} \mid x)$ and both $w_i, w_{i+1} \in [0, 1]$.

For any $y$ and $x \neq x'$ (assuming $x \in [x_i, x_{i+1}]$ and $x' \in [x_j, x_{j+1}]$) that $\mathcal{M}(x) = \mathcal{M}(x') = y$, by our discussions above, that implies there exists $a, a', \tilde{a}, \tilde{a}'$ such that $a + x_i = a' + x_{i+1} = \tilde{a} + x_j = \tilde{a}' + x_{j+1}$. Recall that $p_{A|X}(a \mid x) \leq e^{\epsilon} \cdot p_{A|X}(a' \mid x')$ for all the pairs $(a, x) \neq (a', x')$ (which is enforced by the solver regarding the optimization problem), all conditions in Lemma 3 are satisfied. Therefore, it holds immediately that $p_{Y|X}(y \mid x) \leq e^{\epsilon} \cdot p_{Y|X}(y \mid x')$ for any $x \neq x'$ and the proof is complete. □

## 5.2 Utility Analysis

First, we formally justify that the derived mechanism preserves unbiasedness, which is the key utility of a mean estimator. Although the AAA mechanism is optimized for some specific data distributions, the unbiasedness would not be affected when being applied to any other data distribution, as proven in the following claim.

Claim 2. *Given a perturbation mechanism $\mathcal{M}$ defined as in Section 4.4, it holds that $\mathbb{E}[\mathcal{M}(x)] = x$ for any $x$.*

Here we provide a sketch of the proof. By the definition of $\mathcal{M}$ in Eq. (16), given any $x \in \mathcal{X}$, assuming $x$ falls into an interval $[x_i, x_{i+1}]$, by the formulation of our optimization problem, it is enforced that $\mathbb{E}[A_{x_i}] = \mathbb{E}[A_{x_{i+1}}] = 0$ which implies $\mathbb{E}[A_{x_i} + x_i] = x_i$ and $\mathbb{E}[A_{x_{i+1}} + x_{i+1}] = x_{i+1}$. Also, our definition of $\mathcal{M}$ implies $\mathbb{E}[\mathcal{M}(x)] = w_i(x) \cdot \mathbb{E}[A_{x_i} + x_i](x) + w_{i+1}(x) \mathbb{E}[A_{x_{i+1}} + x_{i+1}]$. Thus,

$$\mathbb{E}[\mathcal{M}(x)] = \frac{x_{i+1} - x}{\sigma} \cdot x_i + \frac{x - x_i}{\sigma} \cdot x_{i+1} = x,$$

and completes the proof.

Next, let's discuss the error. Being consistent with existing notations. Recall $p_X, \hat{p}_X$ to be quantized distribution descriptors of $\mathcal{D}$ with randomized rounding and privacy-preserving mechanisms, and let $p_{A|X}$ be an optimization solution subject to $\hat{p}_X$, which defines the mechanism $\mathcal{M}$. In what follows, let the output variance of $\mathcal{M}$ on $\mathcal{D}$ be $V$ (which is the real performance), and the expected variance subjects to $\hat{p}_X$ be $\hat{V}$ (which is known to us). We define the relative error $\phi \triangleq \frac{\hat{V} - V}{V}$.

Assuming the error for the estimated distribution is bounded (as in Section 3.4), we make the following claim regarding the error on the output variance (which has a direct impact on the utility).

Claim 3. *If it holds that*

$$\left| \frac{p_X(x) - \hat{p}_X(x)}{p_X(x)} \right| \leq \psi \text{ for any } x \in \mathcal{X}, \quad (18)$$

*then we have $|\phi| \leq \min\left\{\frac{\psi}{1+\psi}, \frac{\psi}{1-\psi}\right\}$.*

Proof. Given the estimate $\hat{p}_X$, we obtain an optimized conditional noise distribution $p_{A|X}$ which minimizes the total variance of the output subjects to $\hat{p}_X$, which is defined as

$$\hat{V} = \sum_{i=1}^{N} \hat{p}_X(x_i) \cdot \left( \sum_{a \in \mathcal{A}} a^2 \cdot p_{A|X}(a \mid x_i) \right).$$

As defined in (16), for any sample $x$ from $\mathcal{D}$ (assuming $x \in [x_i, x_{i+1}]$), $\mathcal{M}$ returns values with mean $x_i$ or $x_{i+1}$ in probability. Therefore, it can be shown that the output variance

$$= \sum_{i=1}^{N} p_X(x_i) \cdot \left( \sum_{a \in \mathcal{A}} a^2 \cdot p_{A|X}(a \mid x_i) \right). \quad (19)$$

Please see the full version [49] for detailed derivation. As $\hat{V} > 0$, we have

$$\phi = \frac{\sum_{i=1}^{N} (p_X(x_i) - \hat{p}_X(x_i)) \cdot \sum_{a \in \mathcal{A}} a^2 \cdot p_{A|X}(a \mid x_i)}{\sum_{i=1}^{N} \hat{p}_X(x_i) \cdot \sum_{a \in \mathcal{A}} a^2 \cdot p_{A|X}(a \mid x_i)}. \quad (20)$$

By Eq. (18), it always holds that

$$-\frac{\psi}{1+\psi} \leq \frac{p_X(x) - \hat{p}_X(x)}{\hat{p}_X(x)} \leq \frac{\psi}{1-\psi}. \quad (21)$$

By Eq. (20) and RHS of Eq. (21),

$$\phi \leq \frac{\sum_{i=1}^{N} \frac{\psi}{1-\psi} \cdot \hat{p}_X(x) \cdot \sum_{a \in \mathcal{A}} a^2 \cdot p_{A|X}(a \mid x_i)}{\sum_{i=1}^{N} \hat{p}_X(x_i) \cdot \sum_{a \in \mathcal{A}} a^2 \cdot p_{A|X}(a \mid x_i)} = \frac{\psi}{1-\psi} \quad (22)$$

Similarly, it is easy to show $\phi \leq \psi/(1 + \psi)$, which completes the proof. □

# 6 RELATED WORKS

**Data perturbation under central DP.** The notion of central and local DP are conceptually similar, perturbation algorithms used for central DP mechanisms can often be applied in the contexts of local DP and vice versa, e.g., the Laplace mechanism described in Section 2 plays important roles in the contexts of local and central DP protection. However, the threat models for the central setting are radically different from those for local DP. In the central DP setting, the *trusted* data curator can directly access *all* the original (and sensitive) records, which is not granted in local DP settings. Thus, despite the conceptual similarity, central DP and local DP are metrics in orthogonal contexts and it would be inappropriate to directly compare the central and local differential privacy-enhancing approaches. Unless otherwise stated, the solutions reviewed next do not apply to our target problem setting, and, thus, are orthogonal to this work. In the following, we review several notable fundamental perturbation mechanisms in the central setting that (often) outperform the Laplace mechanism and the Gaussian mechanism.

Ghosh et al. [29] propose the geometric mechanism for enforcing $\epsilon$-DP, for which the distribution of the injected noise is formulated as a two-sided geometric series with tunable parameters and can be viewed as an optimized variant of the Laplace mechanism. In particular, [29] obtains improved result utility under the Bayesian optimization model, in which the privacy practitioner has prior knowledge of the input distribution to calibrate the randomization mechanism. This assumption is fundamentally different from our setting, where no prior knowledge about the input distribution is given.

Another notable series of works that achieve a certain notion of optimality is the staircase mechanisms [26–28]. Specifically, these methods assume that the query (i.e., statistics to be released under DP) is a real-valued function, and the mechanism design aims to minimize the worst-case utility loss for the perturbed output subject to the privacy constraints, which can be formulated as an optimization problem. Specifically, the optimization objective is formulated as a piecewise constant probability density function, which is symmetric to the origin and decreases geometrically, leading to noise distribution with a symmetric staircase shape. It is worth mentioning that the idea of this work is remotely related to the Cactus mechanism [3], which works towards the optimal noise injection mechanism for large-composition, i.e., answering a large number of queries, under the central DP setting through numerical optimization.

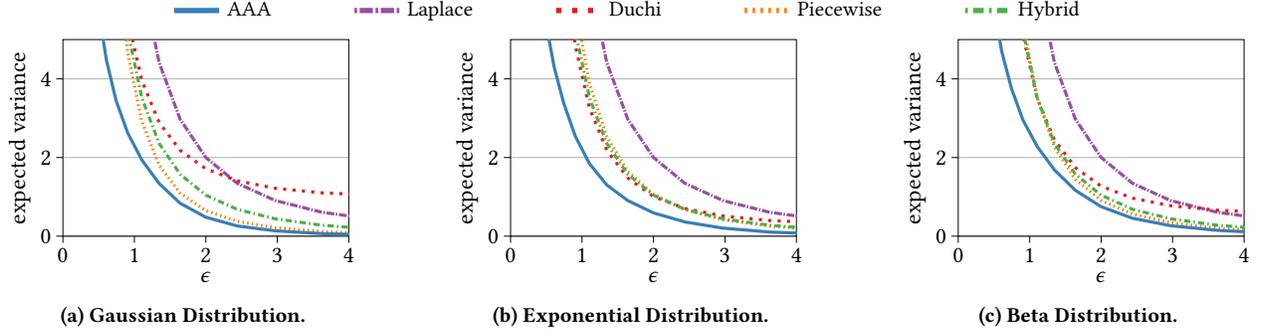

(a) Gaussian Distribution.           (b) Exponential Distribution.           (c) Beta Distribution.

**Figure 3: Comparisons of the expected variance on different input data distributions.**

**The interactive LDP model.** The notion of interactive DP model can be traced back to the classical [20]. There are a few works in the scope of LDP setting [25, 31–33, 40]. Regarding an interactive model, clients respond to the data curator's queries in a sequential manner, and clients are permitted to observe the preceding responses from other clients which affect the perturbation applied to their own data. AAA mechanism falls into this category as we apply a two-step approach, where the responses from a randomly selected portion of clients directly affect the perturbation mechanism applied to the rest of the clients. It has been shown that interactive LDP mechanisms are stronger than the non-interactive ones for tasks such as population quantiles, logistic regression, supporting vector machine, and estimating an unknown Gaussian distribution [25, 31, 32, 40]. On the contrary, the problem of empirical mean estimation for a given set of data points, which is the focus of this paper, has not been investigated under the interactive LDP framework. Our work fills this gap by showing that AAA, which is interactive, achieves better utility than the existing (non-interactive) solutions.

Interactive model is also applied in other contexts of privacy, for example, protecting *Geo-Indistinguishability* [9], an adaptation of differential privacy to the domain of geographic locations. Similar to our proposed mechanism, [9] proposes to formulate an optimization problem with a prior, and the objective is to find a location-randomization mechanism that minimizes the expected distance between original and perturbed locations while preserving geo-indistinguishability. However, it is worth noting that even reducing the problem to one dimension (location-based data are 2-dimensional) and adjusting the constraints according to local DP, the optimized mechanism still fails to serve as a mean estimation mechanism, as the objectives of the problems are different.

## 7 EXPERIMENTS

In this section, we compare the proposed AAA mechanism with several benchmark private mean estimation mechanisms in the literature, i.e., the Laplace mechanism, Duchi's mechanism [19], and piecewise/hybrid mechanisms [44]. In addition, we also compare the squared wave (SW) mechanism [36], as an example that is designed for a different task.

Note that the aforementioned estimators take a similar approach - let local clients perturb the original data, collect the perturbed data, and output the mean. Therefore, the randomness of the perturbation mechanisms would directly impact the final result. Accordingly, our experimental evaluation and comparison of the mechanisms consists of two phases. In Section 7.1, we compare the expected output variances regarding various data distributions, which is the prime target optimization objective of our mechanism. In Section 7.2, we compare the practical performances on various synthetic and real datasets regarding the mean estimation task, which is the ultimate objective of this work.

### 7.1 Evaluating the Expected Variance

In this section, assuming the data distribution $f_X$ is accessible, we compare the expected variance $\mathbb{E}[\text{Var}(Y \mid X)] = \int_{\mathcal{X}} f_X(x) \cdot \int_{\mathcal{Y}} f_{Y|X}(y \mid x) \cdot (y-x)^2 dy dx$. We test the following three types of data distributions: (1) Gaussian distribution $\mathcal{N}(0, 0.1^2)$, which is symmetric to the center of the sample domain (origin). (2) Shifted Exponential distribution, which shifts the distribution function of $\text{Exp}(6)$ to the left by $\beta$, and has probability mass accumulating on one side of the sample domain. (3) Beta distribution $\text{Beta}(0.5, 0.5)$, whose probability mass concentrates on both ends of the domain. For fitting to our setting, all distributions are truncated (with necessary normalization) to the range $[-1, 1]$.

Under the metric of pure-LDP, we vary the privacy parameter $\epsilon$ in the range $(0, 4]$. For AAA mechanism, when solving the discrete optimization problem of Eq. (15), we set the quantization parameter $\sigma = 0.02$ (i.e., $N = 100$), the noise value range parameter $M = k \cdot N$ for $k = 3$ (such that $q_j^{(i)}$ spans between $[-3, 3]$), the geometry series parameter $r = 0.5$. Through the experiments, we find that the choice of $r$ has a negligible impact as long as $r < 0.6$. We will discuss the choice of these parameters in greater detail in Section 7.3.

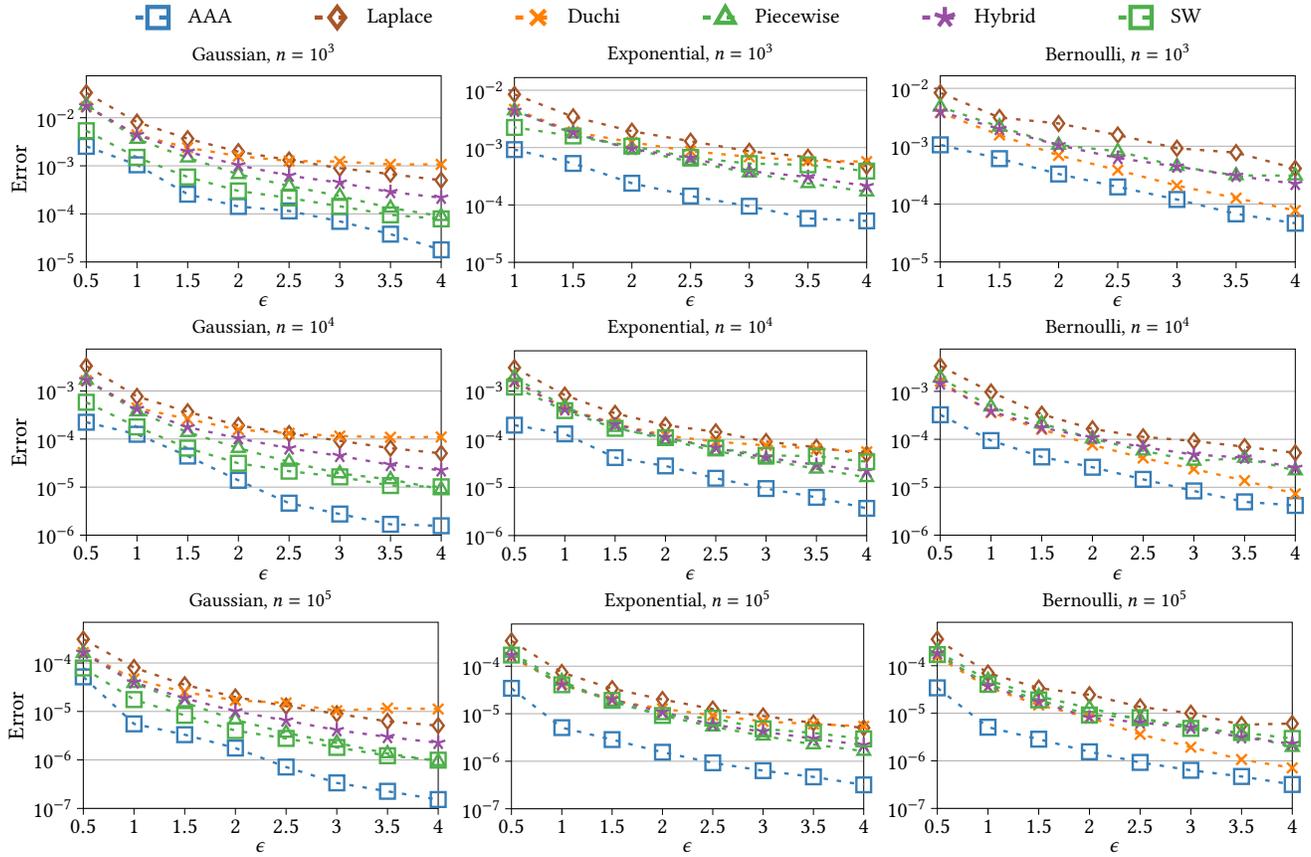

**Figure 4: Performance on synthetic data following different distributions with varying privacy parameters.**

We plot the expected (average) variance as a function of the privacy parameter $\epsilon$ for each of the mechanisms given the ground truth data distributions in Figure 3. Clearly, the proposed AAA mechanism significantly outperforms all of its competitors in all cases. For example, as depicted in Figure 3(a), it is easy to observe that at $\epsilon = 1$, AAA mechanism's expected variance is approximately half of the best result from its competitors. When $\epsilon \to 0$, the performance between AAA and other solutions expands rapidly for all three data distributions.

Another interesting observation from the evaluation results is that although the hybrid mechanism is designed to be an optimized combination of Duchi's mechanism and the piecewise mechanism, its performance under the three data distributions evaluated is sometimes lower than that of the piecewise mechanism, e.g., in Figure 3(a). This confirms our conjecture that high worst-case performance does not necessarily lead to high average-case performance, motivating the usage of our AAA mechanism, which aims to achieve high average-case performance.

## 7.2 Evaluating the Mean Estimation Error

Next, we present the experimental evaluation results with respect to the main task: mean estimation under LDP constraints. We consider the error as the squared difference between the estimated mean and the ground truth mean. That is, let $\mathcal{D}$ be the collection of all local data and $\mathcal{D}_{mean}$ be the $(1 - s)$ portion of the data used for mean estimation. Error $= \left(\sum_{x \in \mathcal{D}_{mean}} \mathcal{M}(x)/|\mathcal{D}_{mean}| - \sum_{x \in \mathcal{D}} x/|\mathcal{D}|\right)^2$. This error is then averaged over 100 independent runs.

We consider synthetic datasets where $n$ samples are drawn from each of the following distributions, with $n \in \{10^3, 10^4, 10^5\}$. Results on higher values of $n$ lead to similar conclusions, and are omitted.

**Gaussian.** Samples from the standard Gaussian distribution $\mathcal{N}(0, 1)$ is are clipped to the range of $[-5, 5]$.

**Exponential.** Samples from the Exponential distribution $\mathrm{Exp}(1)$ are clipped to the range of $[0, 5]$.

**Bernoulli.** Samples from Bernoulli distribution $\mathrm{Ber}(0.5)$ remain.

We also consider three real-world datasets that were used in the previous work [36], described as follows.

**Taxi [42].** This dataset contains the duration of trips (in seconds) from 2018 January New York green taxi data. The range of the data is from 0 to 202989. There are $n = 792744$ samples.

**Income [39].** This dataset contains income information from the 2018 American Community Survey. The range of the income attribute is from $-11200$ to $1423000$. There are $n = 3236107$ samples.

**Retirement [34].** This dataset contains the employee compensation from San Francisco. The range of the data is from $-30621.43$ to $121952.52$. There are $n = 683277$ samples.

For all datasets, we further linearly transform each data point to the range $[-1, 1]$. For estimating the input distribution, we use the generalized randomized response with a portion $s = 0.1$ of the private input data to obtain a noisy histogram, detailed in the full version. Then, we use the estimated input distribution to instantiate AAA mechanism, and inject the additive noise to the remaining $(1 - s)$ portion of the private data. For the AAA mechanism, we set the bin size to 0.125 when $n = 10^5$ and $n = 10^4$ for synthetic data drawn from Gaussian and Exponential as well as for the relatively large-scale real-world datasets while for relatively small synthetic datasets with $n = 10^3$, we set the bin size to 0.25. For synthetic data drawn from the Bernoulli distribution, we choose a larger bin size of 1 as the data domain is only of size 2. We fix $q = \frac{|\mathcal{A}|}{|\mathcal{X}|}$ to 4 for all experiments. More experiments on the hyperparameters are deferred to Section 7.3. We report the average error under different $\epsilon$'s in Figures 4 (for synthetic data) and 5 (for real-world data).

**Synthetic data.** First, we note that AAA consistently achieves the best performance for all synthetic datasets under all privacy constraints. Comparing the performance of AAA across data drawn from different distributions (namely, Gaussian, Exponential, and Bernoulli), we can see that AAA achieves the lowest error on Gaussian, whereas for Exponential and Bernoulli, the error is slightly higher. Intuitively, Gaussian distribution is rather concentrated towards the mean, and hence, it is easier to obtain an accurate histogram estimate by querying only a portion $s$ of the data (especially when $\epsilon$ is large), which, in turn, sets up the stage for optimizing the noise distribution, the key component of AAA. We can also see that for Gaussian data, the improvement of AAA over the baselines becomes higher as $\epsilon$ increases as it is easier to obtain an accurate estimate for the histogram under these settings. On the other hand, it is more difficult to accurately estimate the histogram for the heavy-tailed Exponential. Finally, for the Bernoulli distribution, which has the largest variance itself (namely, error due to sampling), it is naturally more difficult to obtain an accurate mean estimate with a small error, compared with Gaussian and Exponential data.

We next investigate the utility improvement of AAA over the baselines. We can see that the relative improvement achieves the highest for the Gaussian distribution (under large $n$ and $\epsilon$'s) and the Exponential distribution. For Gaussian, as we have explained, it is because the task of histogram estimation is easier. For Exponential, there are more insights. Recall that that $\text{Exp}(1)$ is more skewed toward the left end and as a result, AAA is able to learn "roughly" that the noise distribution should also be skewed toward the left end due to the optimization goal (despite the restrictive privacy constraints). On the other hand, the competitors are not distribution-aware and hence, perform much worse than AAA. Among these competitors, the hybrid mechanism and the piecewise mechanism achieve the lowest error, the reason being that both mechanisms are optimized for the worst-case scenarios (see [44] for more details) and these scenarios happen more often in the heavy-tailed Exponential distributions than the other two.

Regarding the Bernoulli distribution, Duchi's mechanism performs the best among the competitors, and the performance gap between Duchi's and our AAA is less significant, compared with the other two distributions. The reason is that, for Bernoulli data, an intuitively good perturbation strategy is to perturb the data to the boundaries, which resembles Duchi's mechanism. As a result, the improvement of AAA on Bernoulli is not as significant as in the previous two data distributions (Gaussian and Exponential).

Finally, SW performs the best among the competitors for data drawn from the Gaussian distribution, which is the "smoothest" distribution among the three and is naturally in line with the smoothening step of SW to post-process the collected data.

**Real-world data.** In Figure 5, AAA mechanism significantly outperforms its competitors on Green Taxi and Retirement under all privacy constraints. For Income, the improvement of AAA is less significant when $\epsilon$ is large, while the performance of hybrid and piecewise mechanisms is pretty good. The reason is similar to what we have explained above–the Income data is skewed towards the left end while having a heavy tail, which is more favorable to the hybrid and piecewise mechanism, and in the meantime, it is difficult to estimate the histogram accurately for AAA.

## 7.3 Effect of Hyperparameters

There are three hyperparameters in AAA mechanism: split ratio $s$, which quantifies the portion of data used for distribution estimation; the number of bins $N$ that partitions the input data range; and the ratio $q = |\mathcal{A}|/|\mathcal{X}|$, which quantifies the ratio between the value range of the conditional additive noise and the range of the input data. In this subsection, we study the impact of the aforementioned hyperparameters on the utility of AAA, using the synthetic data of varying size $n$ drawn from $\text{Ber}(0.5)$ and $N(0, 1)$, following with the same pre-processing steps as in Section 7.2. Experiments on Exponential data are omitted as they lead to similar conclusions.

**Split ratio.** As depicted in Figure 6(a), We present the error versus varying split ratios from 0.05 to 0.4 while fixing the bin size to 0.125 for Gaussian and 1 for Bernoulli, with $q = 4$. In general, we prefer relatively small values of $s$ (e.g., 0.05 and 0.1). However, we do notice that the performance of AAA is more sensitive to $s$ on Bernoulli-distributed data. Perhaps the presented robustness of AAA is because Gaussian data is more concentrated, such that a larger $s$ gains limited improvement on the sampling error, resulting in similar final performance. However, for the less concentrated Bernoulli, we need more data (i.e. larger $s$) to estimate the mean.

**Bin size.** One may choose a finer quantization level (i.e. smaller bin size) to better capture the characteristics of the data distribution. However, in practice, it may involve a larger relative error for the distribution estimation, which would eventually impact the result utility. In Figure 6(b), we present the error versus varying quantization parameters (interval width) $\sigma$ from 0.125 to 2 while fixing $s = 0.05$ and $q = 4$. As we can see from the figure, for relatively large values of $\epsilon$ (e.g., 4), a small $\sigma$ results in better utility, whereas for moderate $\epsilon$, the impact of $\sigma$ is less significant. This is mainly due to the trade-off between the granularity of distribution estimation versus its error. For small values of $\epsilon$ (e.g., 0.5), the optimal choice of bin size is also larger. Comparing the results between $n = 10^3$ and $n = 10^4$, we notice that the optimal choice of bin size varies as $n$ varies. When $n = 10^3$, the optimal $\sigma$ under $\epsilon = 1$ is 0, 5. Whereas when $n = 10^4$, the optimal choice is around 0.125, which is much smaller. The reason is similar: when $n$ is small, decreasing $\sigma$ does not help with the histogram estimation, but rather, increases the error. Similar conclusions also apply for Bernoulli and Exponential distribuitons. In general, we would recommend moderated choices

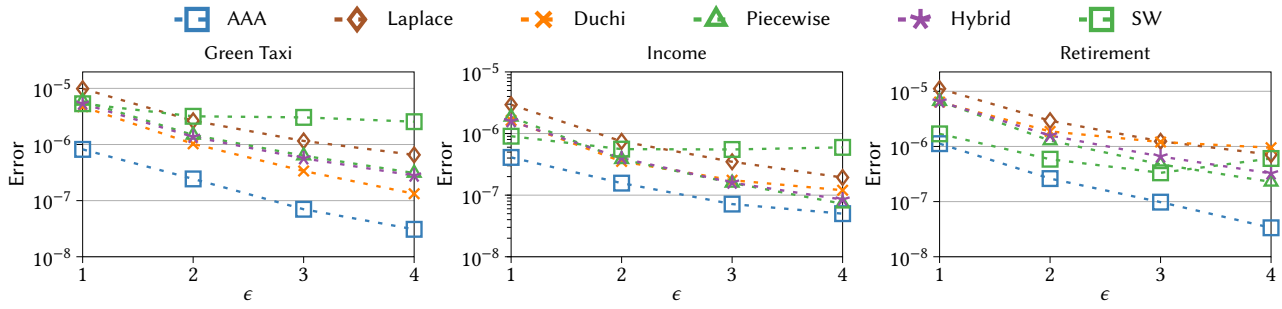

**Figure 5: Performance on real-world datasets under varying privacy parameters.**

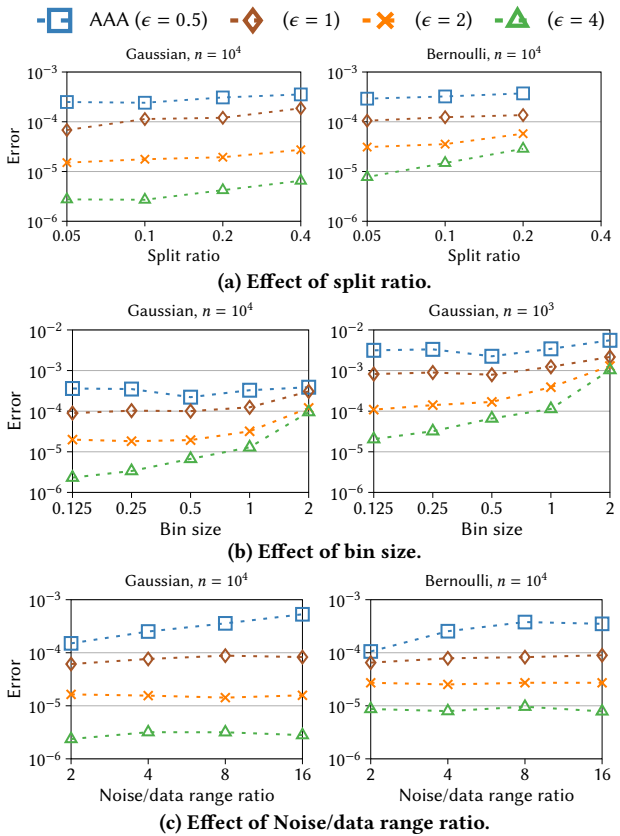

**(a) Effect of split ratio.**

**(b) Effect of bin size.**

**(c) Effect of Noise/data range ratio.**

**Figure 6: Effect of hyperparameters on AAA.**

when either $n$ or $\epsilon$ is small. For large $n$ and $\epsilon$, we suggest relatively small $\sigma$. More results are in the full version [49].

**Noise range.** Parameter $q$ controls the relative range ratio between the noise versus the input data. While a larger value of $q$ requires higher computation costs, it may provide better conditional noise distribution as the solution space is larger (assuming that the estimate for the histogram is accurate). We vary the value of $q = 2, 4, 8, 16$ while fixing the bin size to 0.125 for Gaussian and 1 for Bernoulli, with $s$ to 0.05 and report the results in Figure 6(c). We observe that except for small $\epsilon$ (e.g., 0.5), the impact of $q$ on the utility of AAA is negligible. Overall, we suggest small $q$ to avoid optimizing the noise distribution towards the wrong direction (under restrictively small $\epsilon$) and to save up the computational costs.

## 8 CONCLUSION

In this work, we present a local differentially private mechanism that adapts to the underlying data distribution, whose main component is an algorithm that generates additive noise that is carefully calibrated to achieve an optimal trade-off between privacy guarantee and utility. The proposed AAA mechanism outperforms previous methods in terms of practical performance on common data distributions according to our extensive experiments.

Regarding future work, an interesting direction is to investigate adapting the AAA framework to other data aggregations, e.g. variance estimation, distribution estimation, or federated learning. These tasks require rigorous analysis of the desired characteristics, and we intend to update the AAA mechanism to these more complex LDP applications to obtain improved result utility.

## REFERENCES

[1] Martin Abadi, Andy Chu, Ian Goodfellow, H Brendan McMahan, Ilya Mironov, Kunal Talwar, and Li Zhang. 2016. Deep learning with differential privacy. In *Proceedings of the 2016 ACM SIGSAC conference on computer and communications security*. 308–318.

[2] Jayadev Acharya, Ziteng Sun, and Huanyu Zhang. 2018. Hadamard Response: Estimating Distributions Privately, Efficiently, and with Little Communication. In *AISTATS*.

[3] Wael Alghamdi, Shahab Asoodeh, Flavio P Calmon, Oliver Kosut, Lalitha Sankar, and Fei Wei. 2022. Cactus mechanisms: Optimal differential privacy mechanisms in the large-composition regime. In *2022 IEEE International Symposium on Information Theory (ISIT)*. IEEE, 1838–1843.

[4] Apple. 2016. Differential Privacy Overview. (2016). Retrieved December 21, 2020 from https://www.apple.com/privacy/docs/Differential_Privacy_Overview.pdf

[5] Raef Bassily, Kobbi Nissim, Uri Stemmer, and Abhradeep Guha Thakurta. 2017. Practical Locally Private Heavy Hitters. In *NeurIPS*.

[6] Raef Bassily and Adam Smith. 2015. Local, Private, Efficient Protocols for Succinct Histograms. In *STOC*. 127–135.

[7] Andrea Bittau, Úlfar Erlingsson, Petros Maniatis, Ilya Mironov, Ananth Raghunathan, David Lie, Mitch Rudominer, Ushasree Kode, Julien Tinnés, and Bernhard Seefeld. 2017. Prochlo: Strong Privacy for Analytics in the Crowd. *SOSP* (2017), 441–459.

[8] Keith Bonawitz, Vladimir Ivanov, Ben Kreuter, Antonio Marcedone, H. Brendan McMahan, Sarvar Patel, Daniel Ramage, Aaron Segal, and Karn Seth. 2017. Practical Secure Aggregation for Privacy-Preserving Machine Learning. In *CCS*. 1175–1191.

[9] Nicolás E Bordenabe, Konstantinos Chatzikokolakis, and Catuscia Palamidessi. 2014. Optimal geo-indistinguishable mechanisms for location privacy. In *Proceedings of the 2014 ACM SIGSAC conference on computer and communications security*. 251–262.

[10] Mark Bun, Jelani Nelson, and Uri Stemmer. 2019. Heavy Hitters and the Structure of Local Privacy. *ACM Trans. Algorithms* 15, 4, Article 51 (oct 2019), 40 pages.

[11] Mark Bun and Thomas Steinke. 2019. Average-Case Averages: Private Algorithms for Smooth Sensitivity and Mean Estimation. In *Advances in Neural Information Processing Systems*. 181–191.

[12] Rui Chen, Haoran Li, A. Kai Qin, Shiva Prasad Kasiviswanathan, and Hongxia Jin. 2016. Private spatial data aggregation in the local setting. In *ICDE*. 289–300.

[13] Graham Cormode, Tejas Kulkarni, and Divesh Srivastava. 2018. Marginal Release Under Local Differential Privacy. In *SIGMOD*. 131–146.

[14] Graham Cormode, Tejas Kulkarni, and Divesh Srivastava. 2019. Answering Range Queries under Local Differential Privacy. *PVLDB* 12, 10 (jun 2019), 1126–1138.

[15] Graham Cormode, Samuel Maddock, and Carsten Maple. 2021. Frequency Estimation under Local Differential Privacy. *Proc. VLDB Endow.* 14, 11 (oct 2021), 2046–2058.

[16] Bolin Ding, Janardhan Kulkarni, and Sergey Yekhanin. 2017. Collecting Telemetry Data Privately. In *NeurIPS*. 3574–3583.

[17] Frances Ding, Moritz Hardt, John Miller, and Ludwig Schmidt. 2021. Retiring Adult: New Datasets for Fair Machine Learning. In *NeurIPS*. 6478–6490.

[18] John C. Duchi, Michael I. Jordan, and Martin J. Wainwright. 2013. Local Privacy and Minimax Bounds: Sharp Rates for Probability Estimation. In *NeurIPS*. 1529–1537.

[19] John C Duchi, Michael I Jordan, and Martin J Wainwright. 2018. Minimax optimal procedures for locally private estimation. *J. Amer. Statist. Assoc.* 113, 521 (2018), 182–201.

[20] Cynthia Dwork, Frank McSherry, Kobbi Nissim, and Adam Smith. 2006. Calibrating noise to sensitivity in private data analysis. In *Theory of cryptography conference*. Springer, 265–284.

[21] Úlfar Erlingsson, Vitaly Feldman, Ilya Mironov, Ananth Raghunathan, Kunal Talwar, and Abhradeep Thakurta. 2019. Amplification by Shuffling: From Local to Central Differential Privacy via Anonymity. In *SODA*. 2468–2479.

[22] Úlfar Erlingsson, Vasyl Pihur, and Aleksandra Korolova. 2014. RAPPOR: Randomized Aggregatable Privacy-Preserving Ordinal Response. In *CCS*. 1054–1067.

[23] Alexandre V. Evfimievski, Ramakrishnan Srikant, Rakesh Agrawal, and Johannes Gehrke. 2002. Privacy preserving mining of association rules. In *KDD*. 217–228.

[24] Giulia Fanti, Vasyl Pihur, and Úlfar Erlingsson. 2015. Building a RAPPOR with the Unknown: Privacy-Preserving Learning of Associations and Data Dictionaries. *Proceedings on Privacy Enhancing Technologies* 2016 (03 2015), https://doi.org/10.1515/popets-2016-0015

[25] Marco Gaboardi, Ryan Rogers, and Or Sheffet. 2019. Locally Private Mean Estimation: $Z$-test and Tight Confidence Intervals. In *The 22nd International Conference on Artificial Intelligence and Statistics, AISTATS 2019, 16-18 April 2019, Naha, Okinawa, Japan*. 2545–2554.

[26] Quan Geng, Peter Kairouz, Sewoong Oh, and Pramod Viswanath. 2015. The staircase mechanism in differential privacy. *IEEE Journal of Selected Topics in Signal Processing* 9, 7 (2015), 1176–1184.

[27] Quan Geng and Pramod Viswanath. 2014. The optimal mechanism in differential privacy. In *2014 IEEE International Symposium on Information Theory*. IEEE, 2371–2375.

[28] Quan Geng and Pramod Viswanath. 2015. The optimal noise-adding mechanism in differential privacy. *IEEE Transactions on Information Theory* 62, 2 (2015), 925–951.

[29] Arpita Ghosh, Tim Roughgarden, and Mukund Sundararajan. 2009. Universally utility-maximizing privacy mechanisms. In *Proceedings of the forty-first annual ACM symposium on Theory of computing*. 351–360.

[30] Bernard G. Greenberg, Abdel-Latif A. Abul-Ela, Walt R. Simmons, and Daniel G. Horvitz. 1969. The Unrelated Question Randomized Response Model: Theoretical Framework. *J. Amer. Statist. Assoc.* 64, 326 (1969), 520–539.

[31] Matthew Joseph, Janardhan Kulkarni, Jieming Mao, and Zhiwei Steven Wu. 2019. *Locally Private Gaussian Estimation*.

[32] Matthew Joseph, Jieming Mao, Seth Neel, and Aaron Roth. 2019. The Role of Interactivity in Local Differential Privacy. *2019 IEEE 60th Annual Symposium on Foundations of Computer Science (FOCS)* (2019), 94–105.

[33] Matthew Joseph, Jieming Mao, and Aaron Roth. 2022. Exponential Separations in Local Privacy. *ACM Trans. Algorithms* 18, 4, Article 32 (oct 2022), 17 pages.

[34] Kaggle. 2020. San Francisco Employee Compensation. (2020). Retrieved Feb 21, 2023 from https://www.kaggle.com/datasets/san-francisco/sf-employee-compensation

[35] S. P. Kasiviswanathan, H. K. Lee, K. Nissim, S. Raskhodnikova, and A. Smith. 2008. What Can We Learn Privately?. In *FOCS*. 531–540.

[36] Zitao Li, Tianhao Wang, Milan Lopuhaä-Zwakenberg, Ninghui Li, and Boris Škoric. 2020. Estimating Numerical Distributions under Local Differential Privacy. In *SIGMOD*. 621–635.

[37] Yehuda Lindell and Eran Omri. 2011. A practical application of differential privacy to personalized online advertising. *Cryptology ePrint Archive* (2011).

[38] Zhan Qin, Yin Yang, Ting Yu, Issa Khalil, Xiaokui Xiao, and Kui Ren. 2016. Heavy Hitter Estimation over Set-Valued Data with Local Differential Privacy. In *CCS*. 192–203.

[39] Steven Ruggles, Sarah Flood, Ronald Goeken, Josiah Grover, Erin Meyer, Jose Pacas, and Matthew Sobek. 2019. Integrated Public Use Microdata Series: Version

9.0 [dataset]. (2019). Retrieved Feb 21, 2023 from http://doi.org/10.18128/D010.V9.0

[40] Adam Smith, Abhradeep Thakurta, and Jalaj Upadhyay. 2017. Is Interaction Necessary for Distributed Private Learning?. In *2017 IEEE Symposium on Security and Privacy (SP)*. 58–77. https://doi.org/10.1109/SP.2017.35

[41] Amazon Staff. 2018. Protecting data privacy. https://www.aboutamazon.com/news/amazon-ai/protecting-data-privacy. (2018).

[42] Taxi and Limousine Commission (TLC). 2022. New York Taxi Dataset. (2022). Retrieved Feb 21, 2023 from https://www.nyc.gov/site/tlc/about/tlc-trip-record-data.page

[43] Differential Privacy Team. 2018. Learning with privacy at scale. https://machinelearning.apple.com/research/learning-with-privacy-at-scale. (2018).

[44] Ning Wang, Xiaokui Xiao, Yin Yang, Jun Zhao, Siu Cheung Hui, Hyejin Shin, Junbum Shin, and Ge Yu. 2019. Collecting and analyzing multidimensional data with local differential privacy. In *2019 IEEE 35th International Conference on Data Engineering (ICDE)*. IEEE, 638–649.

[45] Tianhao Wang, Jeremiah Blocki, Ninghui Li, and Somesh Jha. 2017. Locally Differentially Private Protocols for Frequency Estimation. In *USENIX Security*. 729–745.

[46] Tianhao Wang, Bolin Ding, Jingren Zhou, Cheng Hong, Zhicong Huang, Ninghui Li, and Somesh Jha. 2019. Answering Multi-Dimensional Analytical Queries under Local Differential Privacy. In *SIGMOD*. 159–176.

[47] Tianhao Wang, Ninghui Li, and Somesh Jha. 2018. Locally Differentially Private Frequent Itemset Mining. In *2018 IEEE Symposium on Security and Privacy (SP)*. 127–143. https://doi.org/10.1109/SP.2018.00035

[48] Stanley L. Warner. 1965. Randomized Response: A Survey Technique for Eliminating Evasive Answer Bias. *J. Amer. Statist. Assoc.* 60, 309 (1965), 63–69.

[49] Fei Wei, Ergute Bao, Xiaokui Xiao, Yin Yang, and Bolin Ding. 2023. AAA: an Adaptive Mechanism for Locally Differential Private Mean Estimation (Technical report). (2023). Retrieved August 1, 2023 from https://github.com/adaptiveldpmechanism/vldb_2024_submission/blob/main/technical_report.pdf

[50] N.A. Weiss, P.T. Holmes, and M. Hardy. 2006. *A Course in Probability*. Pearson Addison Wesley.

[51] Min Xu, Bolin Ding, Tianhao Wang, and Jingren Zhou. 2020. Collecting and Analyzing Data Jointly from Multiple Services under Local Differential Privacy. *PVLDB* 13, 11 (2020), 2760–2772.

[52] Min Xu, Tianhao Wang, Bolin Ding, Jingren Zhou, Cheng Hong, and Zhicong Huang. 2019. DPSAaS: Multi-Dimensional Data Sharing and Analytics as Services under Local Differential Privacy. *PVLDB* 12, 12 (2019), 1862–1865.

[53] Zhikun Zhang, Tianhao Wang, Ninghui Li, Shibo He, and Jiming Chen. 2018. CALM: Consistent Adaptive Local Marginal for Marginal Release under Local Differential Privacy. In *CCS*. 212–229.

## A  OMITTED CONTENTS

### A.1  Existing Mechanisms for Mean Estimation

To better illustrate the existing mechanisms mentioned in our work, we present them in the form of pseudo codes, the Duchi's mechanism as shown in Algorithm 1, the Piecewise Mechanism in Algorithm 2, and the AAA mechanism in Algorithm 3.

---

**Algorithm 1:** Duchi et al.'s Solution for One-Dimensional Numeric Data [19]

**Input**   :Client data $x \in [-\beta, \beta]$, Local DP parameter $\epsilon$.
**Output**:Perturbed data $y \in \left\{ -\frac{e^\epsilon + 1}{e^\epsilon - 1} \cdot \beta, \frac{e^\epsilon + 1}{e^\epsilon - 1} \cdot \beta \right\}$.

1  Sample a Bernoulli variable $b$ such that
   $\Pr[b = 1] = \frac{e^\epsilon - 1}{2e^\epsilon + 2} \cdot \frac{x}{\beta} + \frac{1}{2}$.
2  **if** $b = 1$ **then**
3  |   $y = \frac{e^\epsilon + 1}{e^\epsilon - 1} \cdot \beta$
4  **else**
5  |   $y = -\frac{e^\epsilon + 1}{e^\epsilon - 1} \cdot \beta$
6  **return** $y$.

---

**Algorithm 2:** Piecewise Mechanism for One-Dimensional Numeric Data [44]

**Input**   :Client data $x \in [-1, 1]$, Local DP parameter $\epsilon$.
**Output**:Perturbed data $y \in [-C, C]$.

1  Sample $\alpha$ uniformly at random from $[0,1]$
2  **if** $\alpha < \frac{e^{\epsilon/2}}{e^{\epsilon/2}+1}$ **then**
3  |   Sample $y$ uniformly at random from $[l(x), r(x)]$
4  **else**
5  |   Sample $y$ uniformly at random from
   |   $[-C, l(x)) \cup (r(x), C]$
6  **return** $y$.

---

**Algorithm 3:** Outline of the Proposed Solution

**Input**   :Collection of client data $\mathcal{D} = \{x_i : i \in [K]\}$, where
        $x_i \sim f_X$; LDP parameter $\epsilon$.
**Output**:A mechanism $\mathcal{M}$ satisfying $\epsilon$-LDP.
**Step 1.** Estimate a quantized data distribution $\hat{p}_X$ with a random subset of the clients under $\epsilon$-LDP, and remove these clients from $\mathcal{D}$.
**Step 2.** Solve the optimization problem in Eq. (15) with respect to the LDP parameter $\epsilon$ using $\hat{p}_X$, to obtain a solution **Q**.
**Step 3.** Use the solution **Q** to compute a discrete conditional distribution $\mathbf{P}_{A|X}$ with Eq. (13), and then obtain a piece-wise constant continuous conditional probability density function $f_{A|X}$ according to Eq. (16).
**Step 4.** Return mechanism $\mathcal{M} : \mathcal{X} \to \mathcal{Y}$ such that $y = \mathcal{M}(x) = a + x$, where $a$ is sampled from the the conditional distribution $f_{A|X=x}$.

---

### A.2  Proof for Eq. (10)

Recall that we aim to investigate $\mathbb{E}[\text{Var}(Y \mid X)]$. More specifically, we have

$$\mathbb{E}[\text{Var}(Y \mid X)] = \mathbb{E}\left[(Y - \mathbb{E}[Y \mid X])^2 \mid X\right]$$

$$= \sum_{x \in \mathcal{X}} \mathbb{E}\left(Y - \mathbb{E}[Y \mid X = x]\right)^2 \cdot p_X(x)$$

$$= \sum_{x \in \mathcal{X}} \left(\mathbb{E}[Y^2 \mid X = x] - \mathbb{E}^2[Y \mid X = x]\right) \cdot p_X(x)$$

$$= \sum_{x \in \mathcal{X}} \left(\sum_{a \in \mathcal{A}} (x + a)^2 \cdot p_{A|X}(a \mid x)da - x^2\right) \cdot p_X(x)$$

$$= \sum_{x \in \mathcal{X}} \left(\sum_{a \in \mathcal{A}} (x^2 + 2xa + a^2) \cdot p_{A|X}(a \mid x) - x^2\right) \cdot p_X(x)$$

$$\overset{(*)}{=} \sum_{x \in \mathcal{X}} \left(\sum_{a \in \mathcal{A}} a^2 \cdot p_{A|X}(a \mid x)\right) \cdot p_X(x).$$

Here $(*)$ holds as by (9) we have

$$\sum_{x \in \mathcal{X}} \left(\sum_{a \in \mathcal{A}} 2xa \cdot p_{A|X}(a \mid x)\right) \cdot p_X(x)$$

$$= \sum_{x \in \mathcal{X}} 2x \cdot \left(\sum_{a \in \mathcal{A}} a \cdot p_{A|X}(a \mid x)\right) \cdot p_X(x) = 0,$$

and $\sum_{a \in \mathcal{A}} p_{A|X}(a \mid x) = 1$ such that

$$\sum_{x \in \mathcal{X}} \left(\sum_{a \in \mathcal{A}} x^2 \cdot p_{A|X}(a \mid x) - x^2\right) \cdot p_X(x)$$

$$= \sum_{x \in \mathcal{X}} \left(x^2 \cdot \sum_{a \in \mathcal{A}} p_{A|X}(a \mid x) - x^2\right) \cdot p_X(x)$$

$$= \sum_{x \in \mathcal{X}} \left(x^2 - x^2\right) \cdot p_X(x) = 0.$$

And that completes the proof for Eq. (10).

### A.3  Proof of Eq. (14)

To complete the proof, the last equality in Eq. (14) is obtained as follows. Let

$$S = \sum_{j=0}^{\infty} (j + M)^2 r^j = M^2 + \sum_{j=1}^{\infty} (j + M)^2 r^j,$$

then

$$rS = \sum_{j=0}^{\infty} (j + M)^2 r^{j+1} = \sum_{j=1}^{\infty} (j + M - 1)^2 r^j,$$

and

$$(1 - r)S = M^2 + \sum_{j=1}^{\infty} (2j + 2M - 1) r^j = (2M - 1) \sum_{j=1}^{\infty} r^j + 2 \sum_{j=1}^{\infty} j \cdot r^j. \quad (23)$$

Again, we play the same trick, let $S' = \sum_{j=1}^{\infty} j r^j = r + \sum_{j=2}^{\infty} j r^j$.

$$S' = \sum_{j=1}^{\infty} j r^j = r + \sum_{j=2}^{\infty} j r^j.$$

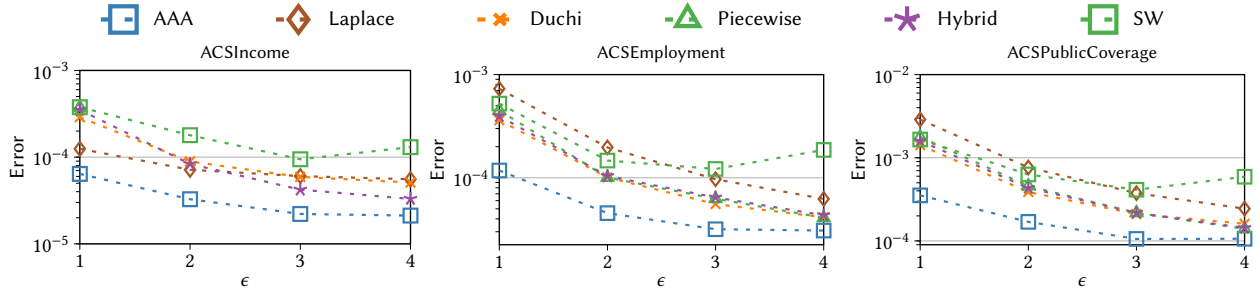

**Figure 7: Performance on multidimensional data under varying privacy parameters.**

and accordingly

$$rS' = \sum_{j=1}^{\infty} jr^{j+1} = \sum_{j=2}^{\infty} (j-1)r^j.$$

So, we have

$$(1-r)S' = r + \sum_{j=2}^{\infty} r^j = r + \frac{r^2}{1-r} = \frac{r}{1-r},$$

thus

$$S' = \frac{r}{(1-r)^2}. \tag{24}$$

Now, plugging (24) into (23), we obtain

$$(1-r)S = M^2 + (2M-1)\frac{r}{1-r} + 2\frac{r}{(1-r)^2}.$$

Namely,

$$S = \frac{M^2}{1-r} + \frac{(2M-1)r}{(1-r)^2} + \frac{2r}{(1-r)^3}.$$

The proof is now complete.

### A.4 Proof of Lemma 3

Without loss of generality, let's assume $a_1 \le a_2$ and $a_3 \le a_4$. For any $w_1 \in [0, 1]$, by linearity it holds that

$$w_1 \cdot a_1 + (1-w_1) \cdot a_2 \le a_2. \tag{25}$$

Similarly, for any $w_2 \in [0, 1]$, we have

$$a_3 \le w_2 \cdot a_3 + (1-w_2) \cdot a_4. \tag{26}$$

By jointly considering our assumption that $a_2 \le k \cdot a_3$ and Eq. (25) and (26), the inequality in the claim hold immediately. We can switch the orders of $\{a_i : i = 1, 2, 3, 4\}$ and it is easy to verify the result still holds and the proof is complete.

### A.5 Proof of Eq. (19)

For computing the output variance on the dataset $\mathcal{D}$, we first need to notice that by the definition of $\mathcal{D}$, assuming $x \in \mathcal{D}$ falls into any interval $[x_i, x_{i+1}]$, it returns randomized values with mean $x_i$ and $x_{i+1}$ in probability. Accordingly, for each index $x_i$, for computing the conditional variance $\text{Var}(Y|X = x) = w_i(x) \cdot \text{Var}[A|X = x_i] + w_{i+1}(x) \cdot \text{Var}[A|X = x_{i+1}]$. Here, $\text{Var}[A|X = x_i] = \sum_{a \in \mathcal{A}} a^2 \cdot$

$\mathbf{p}_{A|X}(a|x_i)$ To compute the overall variance, for each sample $x$ in $\mathcal{D}$, we may consider the probability of $X = x$ is $1/|\mathcal{D}|$. Thus,

$$
\begin{aligned}
V &= \sum_{x \in \mathcal{D}} \sum_{i=1}^{N} \frac{1}{|\mathcal{D}|} \cdot w_i(x) \cdot \text{Var}[A|x_i] \\
&= \sum_{i=1}^{N} \sum_{x \in \mathcal{D}} \frac{w_i(x)}{|\mathcal{D}|} \cdot \text{Var}[A|x_i] \\
&= \sum_{i=1}^{N} \left[ \left( \sum_{x \in \mathcal{D}} \frac{w_i(x)}{|\mathcal{D}|} \right) \cdot \left( \sum_{a \in \mathcal{A}} a^2 \cdot p_{A|X}(a \mid x_i) \right) \right] \\
&\overset{(*)}{=} \sum_{i=1}^{N} p_X(x_i) \cdot \left( \sum_{a \in \mathcal{A}} a^2 \cdot p_{A|X}(a \mid x_i) \right).
\end{aligned}
$$

Here $(*)$ holds by (7).

## B ESTIMATING THE QUANTIZED GLOBAL DATA DISTRIBUTION

As mentioned in Section 3.4, the computation of $\hat{p}_X$ can be done using any existing LDP-compliant histogram estimation algorithm. For completeness, in the following we provide two simple solutions for this purpose, using additive noise and randomized response respectively. We emphasize that these algorithms are *not* our main contribution as the problem has already been well-studied.

### B.1 Using Additive Noise

Algorithm 4 describes a simple LDP-compliant data collection process using additive noise, for the purpose of data distribution estimation.

Using Algorithm 4, the output $y = x + a$ can be viewed as a sample of a random variable $Y = X + A$, where $X$ is the random variable following distribution $f_X$, and $A$ is the injected noise distributed with its own probability density function $f_A$. The probability density function $f_Y$ of $Y$ is then the convolution of $f_X$ and $f_A$, i.e.,

$$f_Y(y) = (f_X * f_A)(y) = \int_{y-\beta}^{y+\beta} f_X(y-\tau) \cdot f_A(\tau) d\tau.$$

The interval of the integration is $[y - \beta, y + \beta]$ as $f_X$ is defined on $\mathcal{X} = [-\beta, \beta]$; thus, $-\beta \le y - \tau \le \beta$, and $\tau \in [y - \beta, y + \beta]$. Therefore, regarding each bin $\mathcal{X}_i$, the quantized estimate $\hat{p}_X$ is $\hat{p}_X(x_i) = \int_{\mathcal{X}_i} f_Y(y) dy$.

## B.2 Using Randomized Response

In this section, we review the classic algorithm for density estimation based on randomized response [30, 48]. Recall from the notations in Section 3.3 that we discretize the original support of the private data $X = [-\beta, \beta]$ to $(N+1)$ edges denoted as $x_0, \ldots, x_N$. These edges form a discrete domain and an induced discrete distribution $P_X$ characterized as in Eq. (7)

We outline the algorithm that computes a private and discrete estimate $\hat{p}_X$ while satisfying $\epsilon$-LDP, as in Algorithm 5. The high-level idea of Algorithm 5 is to let each client randomly perturb her edge index before sending it to the data aggregator, who then reconstructs a histogram based on the perturbed indices.

Here we can use an $N$-by-$N$ matrix $A$ that corresponds to the randomized response Algorithm 5 for histogram estimate. We use the $h$-th row of $A$ (i.e., $A(h)$) to represent the probability distribution that perturbs bin $h$, and the $j$-th column of the $h$-th row (i.e., $A(h, j)$) to represent the probability that the input bin $h$ is mapped to output bin $j$. In particular,

$$A(h, h) = \frac{\exp(\epsilon)}{N + \exp(\epsilon)} \text{ for } h \in \{0, \ldots, N\}; \quad (27)$$

$$A(h, j) = \frac{1}{N + \exp(\epsilon)} \text{ for } j \neq h. \quad (28)$$

It is easy to verify that Algorithm 5 satisfies $\epsilon$-LDP since $A(h, j) \leq \exp(\epsilon) \cdot A(h, j)$, for any $h, h', j \in \{0, \ldots, N\}$. It is also easy to see that matrix $A$ is invertible, and we denote its inverse as $A^{-1}$.

Now we can think of the input private data as a histogram consisting of $(N + 1)$ "bins", written as an $(N + 1)$-dimensional vector $X \in \mathbb{N}^{(N+1)}$. We denote the perturbed histogram collected by the data aggregator as $\tilde{X}$, where each tuple in the bin is perturbed according to the matrix $A$ to preserve LDP. We have that $\mathbb{E}[\tilde{X}] = AX$. To obtain an estimate of $X$, the data aggregator can simply compute

$$\bar{X} = A^{-1}\tilde{X}.$$

It is easy to check that $\mathbb{E}[\bar{X}] = X$ by the linearity of expectation. In practice, we may encounter negative entries in $\bar{X}$, due to the variance introduced by DP noises. If that happens, one can round the negative entries of $\bar{X}$ to 0, and then normalize $\bar{X}$ to obtain a density estimate for the distribution of $P_X$.

## C HANDLING MULTI-DIMENSIONAL DATA

### C.1 Algorithm

---

**Algorithm 4:** Data Collection under LDP with Additive Noise

**Input**   : Client data $x$, LDP parameter $\epsilon$.
**Output**: Perturbed data $y$.
1  Specify a noise $A \sim f_A$ where the parameter of the noise distribution $f_A$ (e.g. Laplace) is computed according to $\epsilon$.
2  Evaluate $A$ to obtain a noise sample $a$.
3  Return $y = x + a$.

---

Regarding the task of collecting multidimensional data for mean estimation under LDP, we apply the methodology mentioned in Section 2.1 that each client proceeds with a random collection of

---

**Algorithm 5:** Density Estimation under LDP with Randomized Response

**Input**   : Collection of client data $\mathcal{D} = \{x_i : i \in [n]\}$, where $x_i \in [-\beta, \beta]$, LDP parameter $\epsilon$, $N + 1$ edges $X_0, \ldots, X_N$.

**Output**: Density estimation $\hat{p}_X$
1  **for** *each client $i$* **do**
2      Client $i$ quantizes her data $x_i$ to one of the $N + 1$ edges, say $b_i$.
3      Client $i$ randomly sample $u_i \sim [0, 1]$.
4      **if** $u_i \geq [0, \frac{\exp(\epsilon)}{N + \exp(\epsilon)}]$ **then**
5         Assign $b_i$ to a uniform sample from $\{0, \ldots, N\}$.
6      Client $i$ sends $b_i$ to the data aggregator.
7  Based on the collected responses $\{b_i\}_i$, the data aggregator constructs a histogram of $N + 1$ edges, denoted as $\tilde{X}$.
8  The data aggregator computes $\bar{X} = A^{-1}\tilde{X}$, where $A^{-1}$ is the inverse of matrix $A$ defined as in Eq. (27) and 28.
9  The data aggregator normalizes $\bar{X}$ to obtain a density estimation $\hat{p}_X$.

---

data entries (attributes) and returns the perturbed version. More specifically, the perturbation is executed in an attribute-wise manner on the individual level, i.e., each client applies the broadcast perturbation mechanism, which is optimized with respect to the estimated distributions, on the assigned attributes. We outline the procedure in Algorithm 6.

### C.2 Experiments

We compare the performance of our solution with the same set of competitors in Section 7 under three real-world datasets, ACSIncome, ACSEmployment, and ACSPublicCoverage [17]. All three datasets are derived from US Census data in year 2018. The number of dimensions $d$ for ACSIncome, ACSEmployment, and ACSPublicCoverage are 11, 17, and 20, respectively. Similar to what we did in Section 7, we normalize each dimension of the datasets to $[-\beta, \beta]$ with $\beta = 1$.

For this set of experiments, each individual client randomly selects $k$ out of $d$ attributes, and privatizes her data on the selected attributes using the perturbation mechanisms, before sending them to the data aggregator, as explained in Section C.1. Without loss of generality, here we set $k = 1$, which does not affect the relative performance of the algorithms. The error is computed as the sum of mean squared errors over all $d$ dimensions. We report the average error over 100 independent runs in Figure 7.

Once again, the AAA mechanism outperforms its competitors by a large margin when $\epsilon$ is small, similar to what we observe in the one-dimensional experiments of Section 7. When $\epsilon$ is large, the improvement is still noticeable although it is not that significant. This suggests that the AAA mechanism can be used as a plug-in replacement for any LDP mechanism for mean estimation, and its improvement is independent of the dimensionality of the data. We do not consider more advanced dimensionality reduction techniques such as [2, 6], as they are orthogonal to the underlying perturbation mechanism, and the argument that AAA

---

**Algorithm 6:** Overview of LDP-compliant multi-dimensional data collection

---

**Input** : Collection of client data $\mathcal{D} = \{\mathbf{x}_i : i \in [n]\}$, where $\mathbf{x}_i = \left(x_i^{(1)}, \ldots, x_i^{(d)}\right) \in [-\beta, \beta]^d$, LDP parameter $\epsilon$, client portion parameter $s$, attribute portion parameter $k$.

**Output** : Mean estimation $\tilde{\mathbf{X}} = \left(\tilde{X}^{(1)}, \ldots, \tilde{X}^{(d)}\right)$

---

**1** **for** *Each client $i$* **do**
**2**    Client $i$ uniformly samples $k$ dimensions from $[d]$ without replacement, obtaining a set of indices denoted as $\mathcal{K}_i$.
**3** **for** *Each attribute $X^{(j)}$* **do**
**4**    $n^{(j)} = \sum_{i=1}^{n} \mathbb{1}(j \in \mathcal{K}_i)$ represents the number of clients that have selected dimension $j$, where the indicator function $\mathbb{1}(j \in \mathcal{K}_i)$ evaluates to 1 if client $i$ has selected dimension $j$ .
**5**    The data aggregator randomly picks a $s$ portion of the $n^{(j)}$ clients that have selected dimension $j$.
**6**    The sampled clients use classic mechanisms satisfying $\epsilon$-local DP to perturb the private attribute $x^{(j)}$.
**7**    The clients return the perturbed attribute $\tilde{x}^{(j)}$ to the data aggregator.
**8**    The data aggregator obtains a distribution estimation $\hat{P}_j$ for the attribute.
**9** The data aggregator obtains AAA mechanisms satisfying $\epsilon$-local DP for each attribute with respect to estimated distribution $\hat{P}_j$.
**10** Broadcast the optimized AAA mechanisms to all clients.
**11** **for** *Each attribute $X^{(j)}$, the rest $1 - s$ portion of clients that have selected dimension $j$* **do**
**12**    The clients use broadcasted AAA mechanisms to perturb the private attribute $x^{(j)}$.
**13**    The clients return the perturbed attribute $\tilde{x}^{(j)}$ to the data aggregator.
**14**    The data aggregator obtains $\tilde{X}^{(j)} = \sum \frac{1}{(1-s)n^{(j)}} \cdot \tilde{x}_i^{(j)}$ as a mean estimation of the attribute.

---

outperforms other existing solutions would still hold when applying these techniques. Similarly, we also omit further experiments on hyperparameters and additional datasets, since the argument still holds as we have shown for one-dimensional data in Section 7.

## D IMPACT OF BIN SIZE ON AAA

We conduct additional study on the impact of the bin size on the performance of AAA for $n = 10^4$ synthetic data points drawn from the Bernoulli distribution Ber(0.5). The split ratio is fixed to $s$ and $q$ is fixed to 4. The result is shown in Figure 8. First, the impact of bin size on AAA under Bernoulli is not as significant as that under Gaussian. In addition, the optimal choice of bin size for AAA under Ber(0.5) tends to be larger than what we have observed for Gaussian data. The reason is that the Bernoulli distribution itself only takes two values and hence, having a finer quantization level does not necessarily lead to better histogram estimation, although the solution space for noise is increased. In particular, for large $\epsilon$'s, the optimal bin size decreases to smaller values whereas for small $\epsilon$'s, we may prefer larger bin sizes.

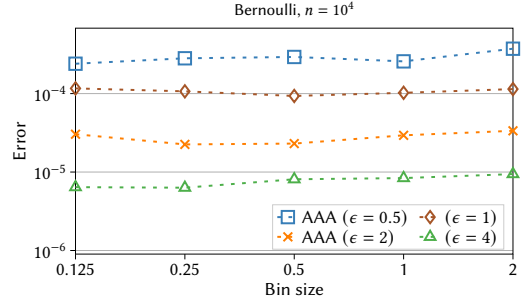

**Figure 8: Impact of bin size on AAA under Ber(0.5).**